\documentclass[
    aps,
    pra,
    twocolumn,
    letterpaper,
    floatfix
]{revtex4-1}
\usepackage{bm,enumerate,dcolumn,tikz,graphicx,color,amsmath,amssymb}
\usepackage{gensymb}
\usepackage{epstopdf}

\newcommand{\overbar}[1]{\mkern 1.5mu\overline{\mkern-1.5mu#1\mkern-1.5mu}\mkern 1.5mu}

\usepackage{hyperref} 
\usepackage{xcolor}
\definecolor{cite}{rgb}{0.,0.,0.85}   
\hypersetup{colorlinks,linkcolor={cite},citecolor={cite},urlcolor={cite}}

\usepackage[normalem]{ulem}
\definecolor{newc}{rgb}{0.,0.,0.5}
\definecolor{newerc}{HTML}{03B4C8}

\begin{document}
\title{ Detecting dark matter waves with a network of precision measurement tools}
\author{Andrei Derevianko}
\affiliation{Department of Physics, University of Nevada, Reno, Nevada
89557, USA}
\begin{abstract}
Virialized Ultra-Light Fields (VULFs) are viable cold dark matter candidates and include scalar and pseudo-scalar bosonic fields, such as axions and dilatons.
Direct searches for VULFs rely on low-energy precision measurement tools.  While the previous proposals have focused  on detecting coherent oscillations of the VULF signals at the VULF Compton frequencies at individual devices, here I
consider a network of such devices.  VULFs are essentially dark matter {\em waves} and as such they carry both temporal and spatial phase information. Thereby, the discovery reach can be improved by using  networks of precision measurement tools. To formalize this idea, I derive a spatio-temporal two-point correlation function for the ultralight dark matter fields in the framework of the standard halo model.  Due to VULFs being Gaussian random fields, the derived two-point correlation function fully determines $N$-point correlation functions. For a network of $N_{D}$ devices within the coherence length of the field, the sensitivity compared to a single device can be improved by a factor of  $\sqrt{N_{D}}$.  Further, I derive a VULF dark matter signal profile for an individual device.  The resulting line shape is strongly asymmetric
due to the parabolic dispersion relation for massive non-relativistic bosons. I  discuss the aliasing effect that extends the discovery reach to VULF frequencies higher than the experimental sampling rate. I present sensitivity estimates and develop a stochastic field SNR statistic. Finally, I consider an application of the developed formalism to atomic clocks and their networks.
\end{abstract}

\maketitle

\section{Introduction}
Exacting the microscopic nature of dark matter (DM) is  one of grand challenges of modern physics and cosmology~\cite{Feng2010,Bertone2005}. 
Here I focus on a general class of DM candidates: ultralight bosonic fields. These include both pseudo-scalar and scalar fields, axions being the example of pseudo-scalars and dilatons -- of the scalars. I will collectively refer to such ultralight fields as VULFs (Virialized Ultra-Light Fields). Such fields behave as  classical entities coherent on a scale of individual devices and can be searched for with  low-energy precision measurements tools.  
Precision measurements, with their exquisite precision, have been
historically important~\cite{Safronova2017-RMP} in powerfully constraining new physics beyond the Standard Model (SM) and can be repurposed for dark matter searches.

 Individual direct VULF DM search proposals cover a broad range of experiments~\cite{ArvHuaTil15,GraKapMar2016,GeraciDerevianko2016-DM.AI,casper,StaFla2015-LaserInterferometry,Arvanitaki2016_SoundDM,Arvanitaki2016-GWdetectors}: atomic clocks, magnetometers, accelerometers, interferometers, cavities, resonators, permanent electric-dipole and parity-violation measurements, and extend to gravitational wave detectors. The sought DM signature in these proposals is DM-induced  oscillations of the measured signals at the VULF Compton frequencies at the device location. 
All these proposals can either discover VULFs or substantially constrain  yet unexplored parameter space.
These distinct tools can be located at geographically separated laboratories
across several continents or in space and one can envision a network of such
tools as a global DM observatory. The network can be heterogenous, i.e., nodes
can be populated with different kinds of devices. 
Here I explore the DM-induced temporal and spatial correlations between nodes of such a global
network with the goal of analyzing the network's discovery potential.

\begin{figure}[!ht]
\begin{center}
\includegraphics[width=0.7\columnwidth]{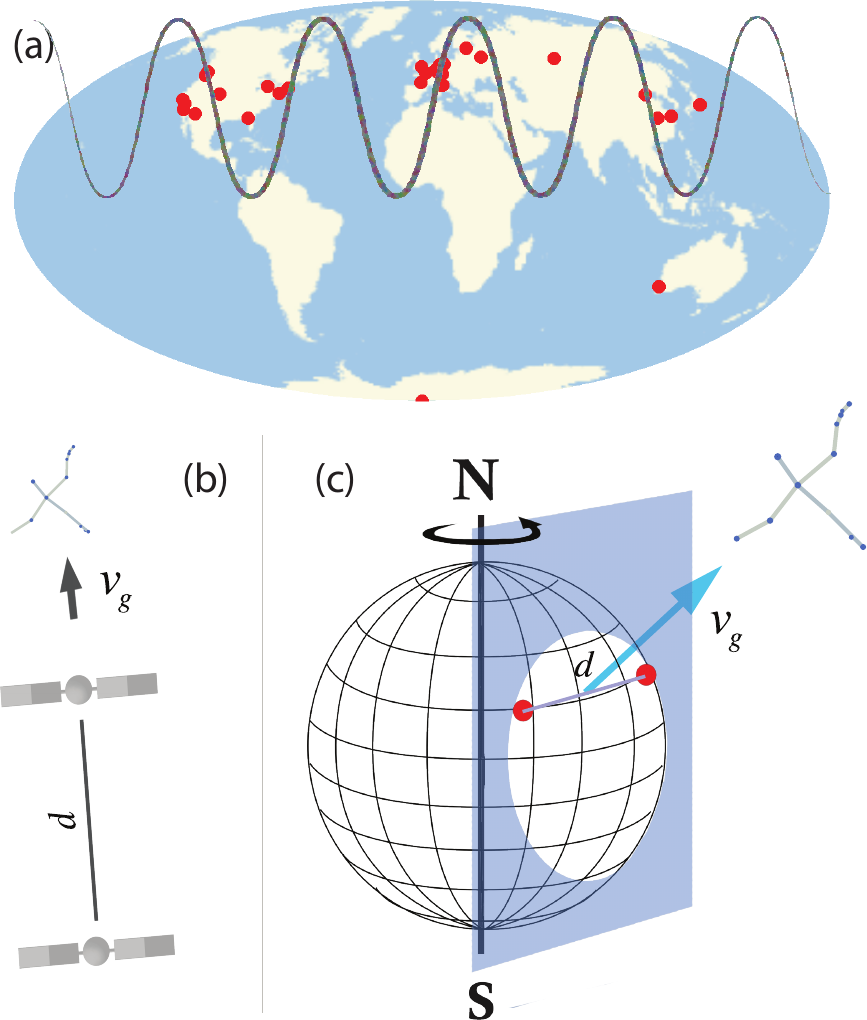}
\caption{ \label{Fig:MapSatsEarth}
(a)  Dark-matter wave observatory based on a global network of existing low-energy precision measurement laboratories (red dots) around the globe;
(b) Satellite mission for probing VULF DM correlation function; both the distance between the satellites and the angle between galactic velocity $\mathbf{v}_g$ and separation $\mathbf{d}$ vectors can be varied.
(c) Terrestrial experiment with fixed nodes utilizes the daily variation of the angle between
galactic velocity and  two-node separation vector.
}
\end{center}
\end{figure}

An important point is that VULFs  are  {\em waves},  and while they  do induce an oscillating in time signal at a given spatial location, DM signals at different locations have a fixed phase relation (see Fig.~\ref{Fig:MapSatsEarth}(a)), i.e., the signals are correlated. Based on this observation, here 
we argue that  a wider discovery reach can be gained by sampling the DM wave at several locations via  a network of precision measurement tools. Further, the VULF signal is composed out of interfering waves traveling at different velocities and in different directions. Then the problem of relating signals at different space-time locations requires computations of dark-matter correlation functions, derived here. Based on these ideas and derivations,  one can envision  a number of  DM wave detection experiments. In the most basic version, the modifications to already running experiments are minor and only require simple GPS-assisted time-stamping of data acquisition~\cite{BudDer2015_PhysicsToday}. Previously, a network of precision measurement devices have been proposed for detecting clumpy DM objects sweeping through the networks~\cite{Pospelov:2012mt,DerPos14}. Here we show that such networks can be also used as discovery tools for VULF ``wavy'' dark matter.

This paper also addresses the difference between deterministic and stochastic nature of ultralight dark matter fields as detected by an individual device. 
It seems that so far the previous VULF literature approach was to treat  ultralight dark matter field signals as deterministic.
This is related to the long coherence times of VULFs on a time scale of a typical measurement campaign.  As discussed in this paper, individual devices
can be also sensitive to VULF  of frequencies much higher than the sampling rate through the aliasing effect inherent to discretized measurements. 
For such high-frequency fields the coherence time is reduced and the more adequate description is the stochastic approach of this paper. I derive the expected line shape of a stochastic dark matter signal and 
present a frequency-space data analysis strategy.   I also discuss an application of the developed formalism to atomic clocks.

This paper is organized as follows. Sec.~\ref{Sec:CorrFunction} presents the derivation of the $N$-point correlation function for ultralight dark matter fields.
Sec.~\ref{Sec:LineShape} presents VULF dark matter line shape for an individual device signal. Sec~\ref{Sec:Network} extends the discussion to a network of devices.
Implications of the formalism and specific data analysis strategies are discussed in Sec~\ref{Sec:Constraints}. This section also addresses the importance of aliasing effect for search for high-frequency VULF DM signals and introduces SNR statistic based on the derived correlation function. An illustrative application of some of the ideas to atomic clocks DM searches is presented in Sec.~\ref{Sec:Clocks}. Finally, Sec~\ref{Sec:Conclusions} draws conclusions. Appendix~\ref{Sec:App:DFT-probabilities} reviews discrete Fourier transform and frequency-space probability distributions for deterministic and stochastic signals. Appendices~\ref{Sec:App:CouplingStrength} and \ref{Sec:App:SumToIntegral} present  derivations of certain equations.
Since the intended audience includes both atomic and particle physics communities,  I restore $\hbar$ and $c$ in the formulas in favor of using natural or atomic units. I use the rationalized Heaviside-Lorentz units for electromagnetism.



\section{Correlation function for ultralight dark matter fields}
\label{Sec:CorrFunction}

In the VULF  models, dark matter is composed of ultralight spin-0 bosonic
fields, oscillating at their Compton frequency $\omega_{\phi}=m_{\phi}%
c^{2}/\hbar$, where $m_{\phi}$ is the boson mass, see e.g., review~\cite{Magana2012}. 
The frequencies can span many orders of magnitude: \ $10^{-10}\,\mathrm{Hz}%
\lesssim f_{\phi}=\omega_{\phi}/\left(  2\pi\right)  \lesssim10^{15}%
\,\mathrm{Hz}$ for $10^{-24}\,\mathrm{eV}\lesssim m_{\phi}\lesssim
10 \,\mathrm{eV}$. Here the lower bound comes from requiring that the virial de Broglie wavelength is smaller than the galactic size and the upper limit --- from requiring that number of particles per de Broglie volume is macroscopic. I formalize these estimates below. 
The proposals~\cite{ArvHuaTil15,GraKapMar2016,GeraciDerevianko2016-DM.AI,casper,StaFla2015-LaserInterferometry,Arvanitaki2016_SoundDM} have focused on searching for an oscillating signal at the Compton frequency.
 Unfortunately, in a laboratory environment, an
observation of an oscillating signal could be ascribed to some mundane ambient
noise and it desirable to establish additional DM signatures. To this end, in this section, I  derive VULF spatio-temporal correlation functions, and explore its experimental significance in later parts of the paper.


\subsection{Linear SM-DM portals}
\label{Sec:Portals}
Additional phenomenological commonality of all the VULF searches is the coupling of DM
fields to SM particles and fields in terms of so-called portals, when the
gauge-invariant operators of the SM fields $O_{X}$ are coupled to the
operators involving DM fields~\cite{Essig:2013lka,DerPos14}. One of possibilities is
the portal linear in the VULF field $\phi\left(  t,\mathbf{r}\right)$,
\begin{equation}
-\mathcal{L}_\mathrm{lin}= \sqrt{\hbar c} \,  \phi\left(  t,\mathbf{r}\right)  \sum_{X} \gamma_{X} \mathcal{O}_{X}. \label{Eq:LinearPortal}
\end{equation}
Here we introduced coupling strengths $\gamma_{X}$; these are to be determined as a result of a positive DM signal detection or constrained otherwise. Fixing the  units of fields $\phi$ to be that of energy, $[\gamma_{X}] = [\mathrm{Energy}]^{-1}$ and one could equivalently parameterize the linear portal~(\ref{Eq:LinearPortal}) in terms of  energy scales $\Lambda_{X} = 1/|\gamma_{X}|$.

For axions and axion-like pseudo-scalar fields (see, e.g., a recent review~\cite{Graham:2015ouw}), the portals are  parameterized as $\sqrt{\hbar c} g_{a\gamma\gamma}\phi \, F_{\mu \nu} \tilde{F}^{\mu\nu}$, $\sqrt{\hbar c} g_{agg} \phi \, G_{\mu \nu} \tilde{G}^{\mu\nu}$, and $\sqrt{\hbar c} g_{aff}  \partial_{\mu} \phi  \, \bar{\psi}_{f} \gamma^{\mu} \gamma^{5} \psi_{f}$, where $F_{\mu \nu}$ and $G_{\mu \nu}$ are Faraday tensors for electromagnetism and QCD, $\tilde F_{\mu \nu}$ and $\tilde G_{\mu \nu}$ are dual tensors,  $\psi_{f}$ are SM fermionic fields,  $\gamma$'s are the Dirac matrices, and $g_{X}$ are coupling constants. Only the last portal does not conform to the  parameterization~(\ref{Eq:LinearPortal}) as it contains the 4-derivative of the axion field. However, when computing action as an integral of the Lagrangian density, the offending 4-derivative $\partial_{\mu} \phi$ can be moved to the fermion current by integrating by parts, thus restoring the canonical parameterization~(\ref{Eq:LinearPortal}).   Further application of the Dirac equation leads to an equivalent form~\cite{Pospelov2008}:
$\partial_{\mu} \phi  \, \bar{\psi}_{f} \gamma^{\mu} \gamma^{5} \psi_{f} \rightarrow 2 m_{f} c^{2 }\, \phi \, \bar{\psi}_{f}  i \gamma^{5} \psi_{f}$,
with $m_{f}$ being the fermion mass. Therefore, the formalism developed in this paper is  applicable to direct searches for axion-like particles.

For scalar fields, such as 
moduli~\cite{Dimopoulos1996,Arkani-Hamed-2000,Burgess2011,Cicoli2011} and dilatons~\cite{Taylor1988,DamourPolyakov1994}
$\mathcal{O}_{X}$ in Eq.~(\ref{Eq:LinearPortal}) are scalars. For example, these could be various pieces from the SM Lagrangian density, $-\mathcal{L}_\mathrm{SM}=\sum_{X} \mathcal{O}_{X}$, such as the fermion rest mass energies $m_{f} c^{2} \bar{\psi}_{f}\psi_{f}$, electromagnetic Faraday tensor contribution $1/4 F_{\mu\nu}F^{\mu\nu}$ , gluon field contribution, etc.  Quite naturally, these portals
when combined with the SM Lagrangian, lead to variation of fundamental
constants, e.g., the electron rest mass $m_{e}$ is modulated by DM field as
$m_{e}\left(  t,\mathbf{r}\right)  = m_{e,0}\times\left(  1+\sqrt{\hbar c} \gamma_{m_{e}} \phi\left(
t,\mathbf{r} \right)\right) $ or the electromagnetic fine structure constant
$\alpha\left(  t,\mathbf{r}\right)  = \alpha_{0}\times\left(  1+\sqrt{\hbar c} \gamma_{\alpha} \phi\left(
t,\mathbf{r}\right) \right) $, where $m_{e,0}$ and $\alpha_{0}$
are unperturbed quantities. VULF fields oscillate at Compton frequencies,
leading to oscillating corrections to fundamental constants. The coupling constants $\gamma_{m_{e}}$ and $\gamma_{\alpha}$ can be expressed in terms of dimensionless dilaton couplings used, e.g., in Refs.~\cite{ArvHuaTil15,Arvanitaki2016_SoundDM}: $d_{e} = \gamma_{\alpha} E_\mathrm{P}/\sqrt{2 \pi}$,
and $d_{m_{e}} =  \gamma_{m_{e}} E_\mathrm{P}/\sqrt{2 \pi}$, where the Planck energy  $E_\mathrm{P} = \sqrt{ \hbar c^{5} /G} \approx 1.22 \times10^{19} \, \mathrm{GeV}$.

A typical apparatus takes  measurements  
associated with an operator $\mathcal{O}_{X}$  (or a combination of operators).
For example, an optical clock measures transition frequencies that depend on
$\alpha$  (and $m_{e}$, see Sec.\ref{Sec:Clocks}.) Then the measured quantity has a DM-induced admixture $
S_{X}\left(  t,\mathbf{r}\right)$ that is proportional to the field value
$\phi\left(  t,\mathbf{r}\right)$ at the device location. Thereby, in the assumption of the linear portal, the correlation between two devices DM signals
can be expressed in terms of the two-point DM field correlation function $g\left(
\tau,\mathbf{d}\right)  =\langle\phi\left(  t^{\prime}=t+\tau,\mathbf{r}%
=\mathbf{r}^{\prime}+\mathbf{d}\right)  \phi\left(  t,\mathbf{r}\right)
\rangle$:
$$
\langle S_{X^{^{\prime}}}\left(  t^{\prime},\mathbf{r}^{\prime}\right)
S_{X}\left(  t,\mathbf{r}\right)  \rangle\propto \gamma_{X} \gamma_{X'} \langle\phi\left(  t^{\prime},\mathbf{r}^{\prime
}\right)  \phi\left(  t,\mathbf{r}\right)  \rangle \,.
$$
Correlation function for spatio-temporal variations of fundamental constants is also expressed in terms of  DM field correlation function, e.g.,
 $$ 
 \frac{\langle \alpha\left(  t',\mathbf{r}'\right) \alpha\left(  t,\mathbf{r}\right)\rangle}{(\alpha_{0})^2} = 
1+ \hbar c \, (\gamma_{\alpha})^{2} \, g\left(
\tau,\mathbf{d}\right)\,.
$$

Derivation of the DM field correlation function $g\left(
\tau,\mathbf{d}\right)$ is the focus of this section.  A limiting case of two co-located devices or of the same apparatus is when
$d=0$. Then the correlation function depends on the delay time $\tau$ and
through the Wiener-Khinchin theorem $g\left(  \tau,0\right)  $ can be
related to the DM-induced frequency spectrum of the device. We will derive the
dark-matter frequency profile (line shape) in Sec.~\ref{Sec:LineShape}.

\subsection{Derivation of 2- and N-point correlation functions}
 Qualitatively, VULF's coherence times and coherence lengths  are related to DM properties. Indeed, in the standard halo
model (see, e.g., Refs.~\cite{Kuhlen2014,Freese2013a}), during the galaxy formation, as DM constituents fall into the
gravitational potential, their velocity distribution in the galactic reference
frame becomes quasi-Maxwellian with a characteristic dispersion (virial) velocity
$v_\mathrm{vir}=\xi~c$, $\xi\approx10^{-3}$ and a cut-off at the galactic escape
velocity. This velocity distribution leads to spectral broadening of
oscillations (dephasing)  characterized by the coherence time $\tau
_{c}\equiv\left(  \xi^{2}~\omega_{\phi}\right)  ^{-1}$. The velocity
distribution also results in a spatial dispersion of individual wave packets,
leading to the coherence length $\lambda_{c}\equiv\hbar/\left(  m_{\phi}\xi
c\right)$. All these coherence properties  formally emerge from the correlation function derived below.

I derive the VULF correlation function $g\left(  \tau,\mathbf{d}\right) $ by generalizing the formalism of
quantum optics to massive spin-0 bosons and quasi-Maxwellian velocity
distribution of DM fields. The major differences with photons are the dispersion relation for massive bosons and the conservation of the total number of particles.  In the following two paragraphs, I use 
natural units for brevity and later restore the fundamental constants. I start the derivation in the galactic reference frame and then transform the result into the moving device frame.
The correlation function is expressed as a trace of field operators and the density matrix $\hat{\rho}$
$$
g\left(  \tau,\mathbf{d}\right)  =\mathrm{tr}\left(
\hat{\rho}\hat{\phi}\left(  t^{\prime},\mathbf{r}^{\prime}\right)  \hat{\phi
}\left(  t,\mathbf{r}\right)  \right). $$
The field operators  are $\hat{\phi}\left(  \underline{x}\right)
=\sum_{\mathbf{k}}\left(  \hat{a}_{\mathbf{k}}e^{-i\underline{k}\cdot \underline{x}}+\hat{a}_{\mathbf{k}}^{\dagger}e^{-i\underline{k}\cdot \underline{x}}\right)
/\sqrt{2V\omega_{k}}$. Here $V$ is the quantization volume, $\underline{k}=(\omega,\mathbf{k})$ and $\underline{x}=(t,\mathbf{r})$ are
4-momentum and 4-position vectors with $\underline{k}\cdot \underline{x}$ denoting their scalar product in flat space-time. $\hat{a}_\mathbf{k}^{\dagger}$ and $\hat{a}_\mathbf{k}$ are bosonic
creation and annihilation operators.  The summation is carried out over the
field modes with frequencies $\omega_{k}=\sqrt{m_{\phi}^{2}+\mathbf{k}^{2}%
}\approx m_{\phi}+\mathbf{k}^{2}/2m_{\phi}$.  The density matrix is defined as $\hat \rho = \sum_{\{n_{k} \} } P( \{n_{k} \}) |\{n_{k} \} \rangle \langle \{n_{k} \}|$, where Fock states are $|\{n_\mathbf{k} \} \rangle = | n_{1}, n_{2},\ldots \rangle$, with $n_{k}$ specifying occupation numbers of mode $\mathbf{k}$ and $P( \{n_\mathbf{k} \})= \prod_\mathbf{k} P(n_\mathbf{k})$ being the probability of finding the ensemble in a particular Fock state.
$P(n_\mathbf{k}) = f_\mathrm{DM,k}\left(  \mathbf{k}\right) (2\pi)^{3}/V$, where $f_\mathrm{DM,k}\left(  \mathbf{k}\right)$ is the DM momentum distribution normalized with respect to $\mathbf{k}$. 

Taking into account
that the average mode occupation numbers $\bar{n}_\mathbf{k}$ are macroscopic and taking the continuous limit, 
$$g\left(
\tau,\mathbf{d}\right)  \approx\frac{1}{\left(  2\pi\right)  ^{3}}\int
d^{3}k\frac{1}{\omega_{k}}\bar{n}_\mathbf{k}\cos\left(  \underline{k} \cdot \left(  \underline{x}-\underline{x}^{\prime
}\right)  \right) \,.$$ 
Here  the average mode occupation numbers $\bar{n}_\mathbf{k}$ are related to the
 DM energy density $\rho_\mathrm{DM}$ and DM momentum distribution
 as $\bar{n}_\mathbf{k}=\left(  2\pi\right)
^{3}\rho_\mathrm{DM}~f_\mathrm{DM,k}\left(  \mathbf{k}\right)  /m_{\phi}$. Here I used the fact that for non-relativistic particles, number density is $\rho_\mathrm{DM}/m_{\phi}$. The requirement that the occupation
numbers are macroscopic, $\bar{n}_\mathbf{k} \gg 1$, leads
to $m_\phi \ll 10 \,\mathrm{eV}$. 

The resulting two-point correlation function reads (restoring fundamental constants)
\begin{align}
g\left(  \tau,\mathbf{d}\right)  = & \left(  \frac{\hbar}{m_{\phi}c}\right)
^{2}\rho_\mathrm{DM} \int d^{3}v~\frac{f_\mathrm{DM}\left(  \mathbf{v}\right)  }{1+\frac
{1}{2}\left(  \mathbf{v}/c\right)  ^{2}} \times  \nonumber\\
& \cos\left(  \frac{m_{\phi}c^{2}}%
{\hbar}\tau-\frac{m_{\phi}\mathbf{v}}{\hbar}\cdot\mathbf{d+}\frac{m_{\phi
}\mathbf{v}^{2}}{2\hbar}\tau\right)  .
\end{align}
Given the DM velocity distribution~\cite{Vogelsberger2009}, $f_\mathrm{DM}\left(  \mathbf{v}\right)$, this
expression can be evaluated numerically.

Analytical result can be obtained by
taking the  Maxwellian distribution of the standard halo model, $f_\mathrm{DM}\left(
\mathbf{v}\right)  =\left(  2\pi\right)  ^{-3/2}\left(  \xi c\right)
^{-3}\exp\left(  -\frac{\left(  \mathbf{v}-\mathbf{v}_{g}\right)  ^{2}%
}{2\left(  \xi c\right)  ^{2}}\right)  $, where $\xi c$ is the virial velocity
and $\mathbf{v}_{g}\approx10^{-3}c$ is the Earth's velocity in the
galactic reference frame. We further take the galactic escape velocity cutoff
to be infinite and neglect the non-relativistic kinetic energy correction in
the denominator.  The resulting correlation function reads%
\begin{equation}
g\left(  \tau,\mathbf{d}\right)  \approx \frac{1}{2}\Phi_{0}^{2}~\mathcal{A}\left(
\tau,\mathbf{d}\right)  \cos\left(  \omega_{\phi}^{\prime}\tau-\mathbf{k}%
_{g}\cdot\mathbf{d}+\Psi\left(  \tau,\mathbf{d}\right)  \right)  . 
\label{Eq:FieldCorrFuncAnalyl}
\end{equation}
Here $\omega_{\phi}^{\prime}$ is the Doppler-shifted value of the Compton
frequency $\omega_{\phi}^{\prime}=\omega_{\phi}+m_{\phi}v_{g}^{2}/(2\hbar)$
and $\mathbf{k}_{g}=m_{\phi}\mathbf{v}_{g}/\hbar$ is the ``galactic''
wave vector associated with the apparatus motion through the dark matter halo. The
effective field amplitude $\Phi_{0}$ is related to DM energy density as $\Phi_{0}=\frac{\hbar}{m_{\phi}c}\sqrt{2\rho_\mathrm{DM}}$, which comes from directly evaluating the temporal $(00)$ component of the stress-energy tensor for the bosonic field. Correlation amplitude $\mathcal{A}\left(
\tau,\mathbf{d}\right)$ and phase $\Psi\left(  \tau,\mathbf{d}\right)  $ are
defined as
\begin{align}
\mathcal{A}\left(  \tau,\mathbf{d}\right)   &  =\frac{\exp\left(  -\frac{\left\vert
\mathbf{d}-\mathbf{v}_{g}\tau\right\vert ^{2}}{2\lambda_{c}^{2}}\frac
{1}{1+(\tau/\tau_{c})^{2}}\right)  }{\left(  1+(\tau/\tau_{c})^{2}\right)
^{3/4}}, \\
\Psi\left(  \tau,\mathbf{d}\right)   &  =-\frac{\left\vert \mathbf{d}%
-\mathbf{v}_{g}\tau\right\vert ^{2}}{2\lambda_{c}^{2}}\frac{\tau/\tau
_{c}}{1+(\tau/\tau_{c})^{2}}+\frac{3}{2}\tan^{-1}\left(  \tau/\tau_{c}\right) \nonumber
,
\end{align}
where the coherence time $\tau_{c}\equiv\left(  \xi^{2}~\omega_{\phi}\right)
^{-1}\approx10^{6}/\omega_{\phi}$ and length $\lambda_{c}\equiv\hbar/\left(
m_{\phi}\xi c\right)  $ are expressed in terms of the virial velocity $\xi
c\approx10^{-3}c$. The correlation function encodes the priors on VULFs
and DM halo, such as the DM energy density in the vicinity of the Solar
system~\cite{Catena2010}, $\rho_\mathrm{DM}\approx0.3~\mathrm{GeV/cm}^{3}$, motion through the DM halo
at $\mathbf{v}_{g}$ and the virial velocity $\xi c$. Thereby, the correlation function provides
an improved statistical confidence in the event of an observation of DM signal.

The derived correlation function is a {\em two-point} correlation function, while, in general, a network could have several nodes. The $N$-point correlation function is expressed in terms of two-point correlation functions. Indeed, the  field is 
composed from a macroscopic number of individual waves (see field operators $\hat{\phi}(\underline{x})$), and, due to the central-limit theorem, the resulting field is Gaussian in nature (see, e.g, Sec.~16.3 of Ref.~\cite{Peacock-cosmologicalPhysics-book-1999}). For Gaussian random fields, the {\em $N$-point}  correlation function is fully expressed in terms of the derived  {\em two-point} correlation function, see, e.g., appendix E of Ref.~\cite{weinberg-cosmology-2008}. The $N$-point correlation function vanishes for odd $N$ and for even $N$ is expressed as a sum of all possible products of pair-wise two-point correlation functions. For example, for $N_{d}=4$ nodes, 
$\langle \phi(\underline{x}_{1}) \phi(\underline{x}_{2}) \phi(\underline{x}_{3}) \phi(\underline{x}_{4}) \rangle = 
g(\underline{x}_{1},\underline{x}_{2})g(\underline{x}_{3},\underline{x}_{3}) + 
g(\underline{x}_{1},\underline{x}_{3})g(\underline{x}_{2},\underline{x}_{4}) + 
g(\underline{x}_{1},\underline{x}_{4})g(\underline{x}_{2},\underline{x}_{3})
$, where each of two-point correlation functions $g(\underline{x}_{i},\underline{x}_{j}) \equiv g(\tau=t_{i}-t_{j}, \mathbf{d}= \mathbf{r}_{i}-\mathbf{r}_{j})$ is given by Eq.~(\ref{Eq:FieldCorrFuncAnalyl}).

Finally, in the limit when both the coherence length $\lambda_{c}$ and time $\tau_{c}$
are infinitely large, one recovers the fully coherent wave correlation function,
\begin{equation}
g_{\mathrm{coh}}\left(  \tau,\mathbf{d}\right)  =\frac{1}{2}\Phi_{0}%
^{2}~\cos\left(  \omega'_{\phi}
\tau-\mathbf{k}_{g}\cdot\mathbf{d}\right).
\label{Eq:gCoherent}
\end{equation}

\section{VULF parameter space}
\label{Sec:Discussion}

Now with the correlation function at hand, first we explore the vast VULF parameter
landscape. The  conversion formulas of VULF masses to Compton frequency,  coherence time, and coherence length are
\begin{eqnarray}
f_{\phi} &=& 2.42 \times 10^{5} \left( \frac{m_{\phi} c^{2}}{\mathrm{neV}} \right)  \, \mathrm{Hz} \,,\\ 
\tau_{c} &=& 1.59 \times 10^{5} \left(\frac{\mathrm{Hz}}{f_{\phi}}\right)\, \mathrm{s}  \,, \label{Eq:PracticalTauC}\\ 
\lambda_{c} &=& 4.77 \times 10^{7} \left(\frac{\mathrm{Hz}}{f_{\phi}}\right)\, \mathrm{km} \label{Eq:CohLengthPractical}\,.
\end{eqnarray}  
Typical values of these parameters are compiled in Table~\ref{Tab:VULF-params}.
Coherence time is roughly $10^{5}$ of Compton periods. The coherence 
length can be interpreted as the de Broglie wavelength of a particle moving at the
virial velocity and it is  a thousand times larger than the Compton
wavelength. The number density of VULF particles $\rho_\mathrm{DM}/\left(  m_{\phi
}c^{2}\right)  $ ranges from $10^{8}$ to $10^{32}\,\mathrm{cm}^{-3}$
for the indicated masses in Table~\ref{Tab:VULF-params} , i.e.,  a typical device interacts with a macroscopic
number of DM particles. Compton frequencies range from nHz to PHz. Notice that
one oscillation per year corresponds to $f_{\phi}=3\times10^{-8}~\mathrm{Hz}$
(for $m_{\phi}\sim10^{-22}~\mathrm{eV}$). As points of reference for the
coherence length, the size of our galaxy is $\sim10^{18}~\mathrm{km}$ and
the Earth diameter $\sim10^{4}~\mathrm{km}$.

\begin{table}[ht]
\begin{center}%
\begin{tabular}[c]{llll}%
\hline\hline
$m_{\phi},\mathrm{eV}$ & $f_{\phi},\mathrm{Hz}$ & $\tau_{c},\mathrm{s}$ &
\multicolumn{1}{c}{$\lambda_{c}~\&~k_{g}^{-1},\mathrm{km}$}\\
\hline
$10^{-24}$ & $2\times10^{-10}$ & $7\times10^{14}$ & $2\times10^{17}$\\
$10^{-20}$ & $2\times10^{-6}$ & $7\times10^{10}$ & $2\times10^{13}$\\
$10^{-15}$ & $2\times10^{-1}$ & $7\times10^{5}$ & $2\times10^{8}$\\
$10^{-10}$ & $2\times10^{4}$ & $7\times10^{0}$ & $2\times10^{3}$\\
$10^{-5}$ & $2\times10^{9}$ & $7\times10^{-5}$ & $2\times10^{-2}$\\
$1$ & $2\times10^{14}$ & $7\times10^{-10}$ & $2\times10^{-7}$\\
\hline \hline
\end{tabular}
\caption{\label{Tab:VULF-params} Parameters of VULF dark matter for a range of masses $m_\phi$: Compton frequency $f_\phi$, 
coherence time $\tau_c$ and length $\lambda_c$, and the inverse galactic wave-vector $k_{g}^{-1} \sim \lambda_c$ associated with the Solar system motion through the DM halo.}
\end{center}
\end{table}

\section{Dark matter line shape}
\label{Sec:LineShape}

For a single geographic location, the associated power spectral density (PSD)
in frequency space of a coherent signal is a spike at the Doppler-shifted Compton frequency. 
However, for a stochastic field, the PSD is distributed over a range of frequencies. In this section, I derive the relevant  DM-induced spectral line shape. Such spectral profile can be used as a DM signature. 

For a single apparatus, the ``local'' temporal
correlation function reads
\begin{align}
g\left(  \tau,\mathbf{d}=\mathbf{0}\right) = &\frac{1}{2}\Phi_{0}^{2}~\frac{\exp\left(
-\frac{1}{2}\left(  \frac{v_{g}\tau}{\lambda_{c}}\right)  ^{2}\frac
{1}{1+(\tau/\tau_{c})^{2}}\right)  }{\left(  1+(\tau/\tau_{c})^{2}\right)
^{3/4}} \times \nonumber \\
& \cos\left(   \omega'_{\phi}
\tau+\Psi\left(  \tau,\mathbf{0}\right)  \right)  .\label{Eq:LocalCorrelation}%
\end{align}
Notice the presence of the coherence length in the combination $v_{g
}\tau/\lambda_{c}$; it arises due to our motion through the DM halo over the
lag time $\tau$  sampling DM fields $v_{g}\tau
\approx10^{-3}c\tau$ distance apart. For $\tau=1\,\mathrm{s}$ this translates
into a $\sim 300~\mathrm{km}$ distance. Considering that $v_{g}\sim
v_\mathrm{vir},~v_{g}\tau/\lambda_{c}\sim\tau/\tau_{c}$. The signal primary
oscillation frequency $ \omega'_{\phi}$ also depends on the device galactic velocity $v_{g}$ through the DM halo via the Doppler shift for massive particles. Since the Earth velocity
changes seasonally, annual velocity modulations are imprinted in the
correlation function. 

In practice, one could obtain a time series of measurements $\{d_n\}$ at $t_n=n\, t_0$, compute auto-correlation function $G_k = \langle d_n d_{n+k} \rangle$ and see if it fits Eq.~(\ref{Eq:LocalCorrelation}). 
More practical approach, especially for devices exhibiting colored noise, is to work in the frequency space and examine the power spectral density of  ${d_n}$ given by the Fourier transform of $G_k$. This strategy is formalized in Sec.~\ref{Sec:Constraints}.

\begin{figure}[!t]
\begin{center}
\includegraphics[width=0.9\columnwidth]{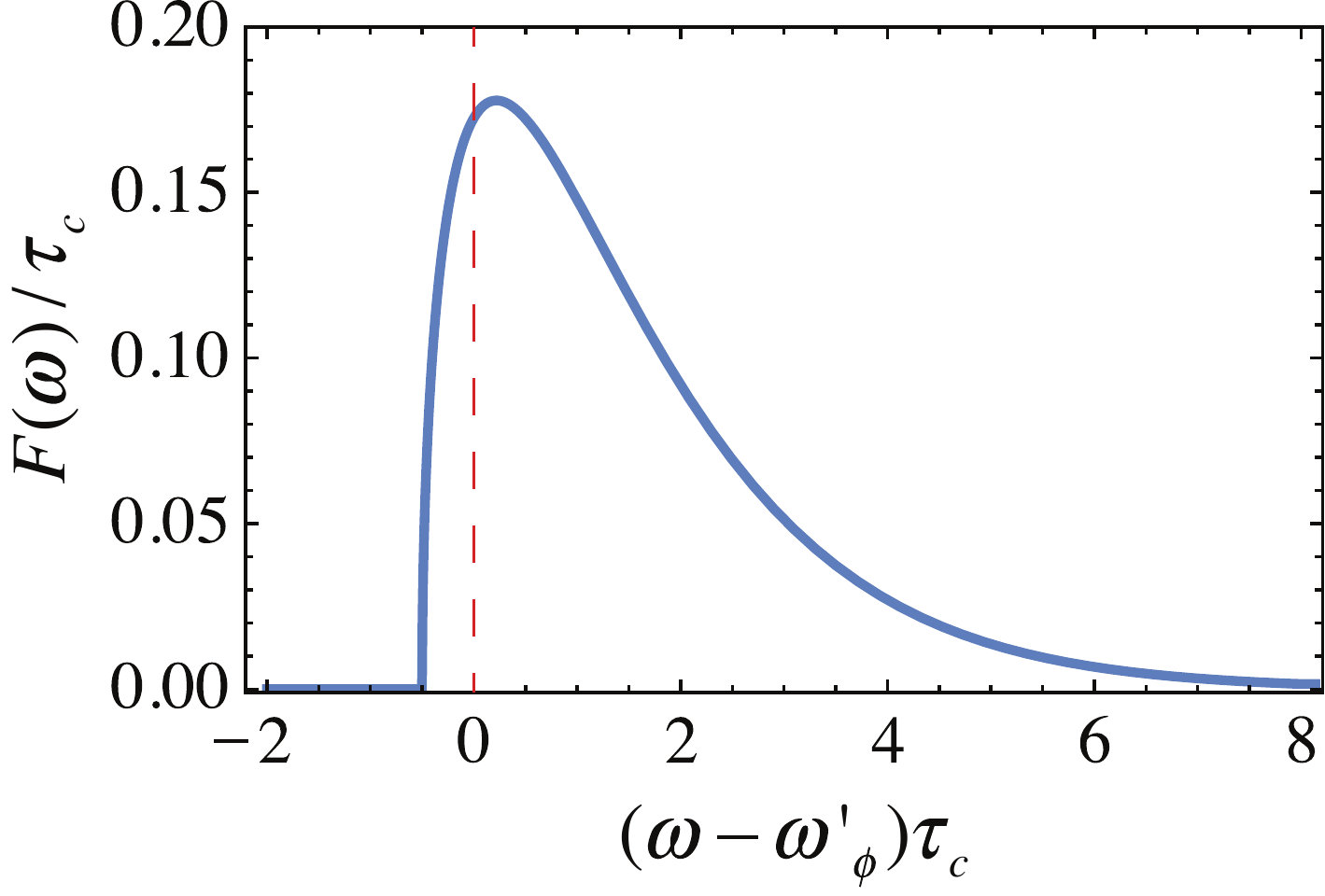}
\caption{ VULF dark matter line profile for co-located  linear portal-sensitive devices or individual apparatus for $\eta=v_g/v_\mathrm{vir}=1$. Dashed vertical line marks the position of the 
Doppler-shifted Compton frequency $\omega_{\phi}^\prime =  \omega_{\phi}+m_{\phi}v_{g}^{2}/(2\hbar)$. The profile value is strictly zero for $\omega < \omega_{\phi}^{\prime}-\eta^{2}/\left(
2\tau_{c}\right)$ due to the dispersion relation for massive particles. The maximum  value of $F(\omega) \approx0.18\tau_{c}$ is attained at  $\omega
\approx\omega_{\phi}^{\prime}+0.22/\tau_{c}$ and the width at half-maximum
is $\Delta \omega_\phi \approx2.5/\tau_{c}$. 
\label{Fig:DMprofile}}
\end{center}
\end{figure}

To facilitate the frequency-space data analysis,  I define the {\em DM-induced line-shape} as a Fourier transform of the
correlation function $f\left(  \omega\right)  =\frac{1}{2\pi}\int_{-\infty
}^{\infty}d\tau~g\left(  \tau,d=0\right)  e^{i \omega\tau}=\frac{1}{2}%
\Phi_{0}^{2}~F\left(  \omega\right)$. $F\left(  \omega\right)  $ is
normalized as  $\int_{0}^{\infty}F\left(  \omega\right)  d\omega=1/2$. With
$\eta=v_{g}/v_{\mathrm{vir}}=v_{g}/\left(  \xi~c\right)$
and the Doppler-shifted frequency $\omega_{\phi}^{\prime}=\omega_{\phi}+m_{\phi}v_{g}^{2}/2$, the
resulting ``dark matter line-shape'' reads
\begin{align}
F\left(  \omega\right)  =  &\left(  2\pi\right)  ^{-1/2}\tau_{c}~\eta
^{-1}  e^{-\eta^{2}}e^{-\left(  \omega-\omega_{\phi}^{\prime}\right)  \tau_{c}%
} \times \nonumber \\ 
& \sinh\left(  \eta\sqrt{\eta^{2}+2\left(  \omega-\omega_{\phi}^{\prime
}\right)  \tau_{c}}\right)  .\label{Eq:DMLineProfile}%
\end{align}
This expression holds for detunings $\omega-\omega_{\phi}^{\prime}>-\eta^{2}/\left(
2\tau_{c}\right)  $, otherwise $F\left(  \omega\right)  =0$. The profile is
shown in Fig.~\ref{Fig:DMprofile}. As expected, the line-width is $\sim\tau_{c}^{-1}$. The
profile is strongly asymmetric due to the parabolic dispersion relation for massive non-relativistic bosons: 
$\omega \approx m_{\phi}c^{2}/\hbar+m_{\phi}\mathbf{v}^{2}/(2\hbar)$. The frequency is shifted to the blue for any finite value of velocity $\mathbf{v}$.   For the fiducial value of $\eta=1$, the maximum $F\left(
\omega\right)  $ value of $\approx0.18\tau_{c}$ is attained at  $\omega
\approx\omega_{\phi}^{\prime}+0.22/\tau_{c}$ and the width at half-maximum
is $\Delta \omega_\phi \approx2.5/\tau_{c}$. 
The DM line shape is  broad in the spectroscopic sense as $\Delta \omega_\phi /\omega_\phi  = 3 \times 10^{-6}$. 

\section{Network}
\label{Sec:Network}
\begin{figure}[!t]
\begin{center}
\includegraphics[width=0.9\columnwidth]{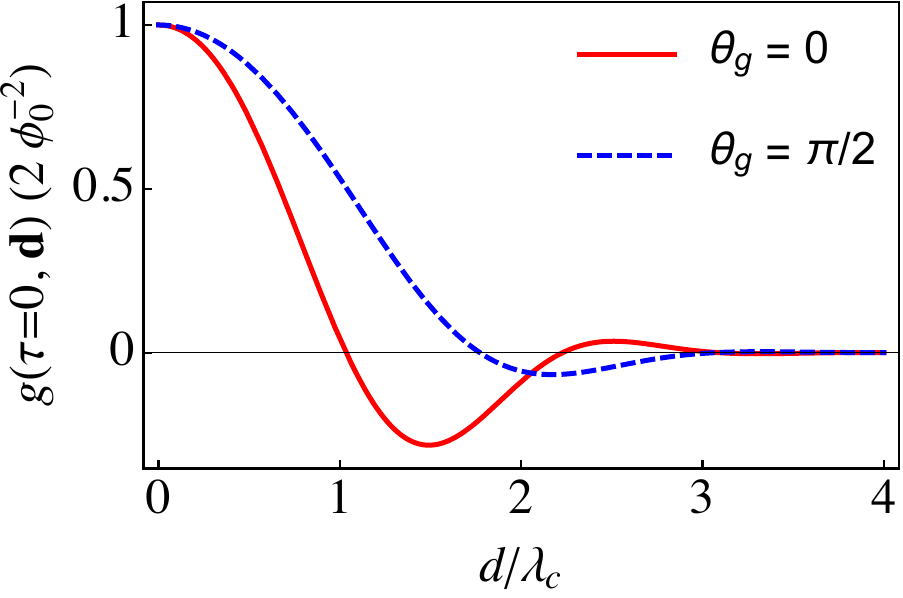}
\caption{ Spatial dependence of VULF two-point correlation function for $\eta=v_g/v_\mathrm{vir}=1$. The red solid curve illustrates the dependence for the case when device separation vector $\mathbf{d}$ is aligned with the galactic velocity $\mathbf{v}_{g}$ and the dashed blue curve --- when $\mathbf{d} \perp \mathbf{v}_{g}$. The galactic orientation angle is defined as $\theta_{g} = \cos^{-1} ( \mathbf{k}_{g} \cdot \mathbf{d})$.
\label{Fig:gSpatialCorrelation}}
\end{center}
\end{figure}

To start the discussion, consider two spatially separated nodes with simultaneously taken measurements, so that the two-point correlation function~(\ref{Eq:FieldCorrFuncAnalyl}) is evaluated at the time lag $\tau=0$.  Fig.~\ref{Fig:gSpatialCorrelation} presents the spatial dependence of such correlation function for two devices. The spatial dependence is determined by two factors, the correlation length $\lambda_{c}$ and the dot product $\mathbf{k}_{g}\cdot\mathbf{d}=m_{\phi}\mathbf{v}_{g}\cdot
\mathbf{d}/\hbar$. Because the galactic wave vector is $k_{g} = \eta/\lambda_{c}\approx 1/\lambda_{c}$, the correlation function in Fig.~\ref{Fig:gSpatialCorrelation} is strongly attenuated, exhibiting only half of an oscillation. This behavior is to be contrasted with the fully coherent correlation function~(\ref{Eq:gCoherent}). The attenuation is maximal when the device separation vector $\mathbf{d}$ is orthogonal to the galactic velocity $\mathbf{v}_{g}$, see blue dashed curve in Fig.~\ref{Fig:gSpatialCorrelation}. In this case, the minor oscillating behavior is entirely due to the phase $\Psi\left(  \tau=0,\mathbf{d}\right)$ in the two-point correlation function~(\ref{Eq:FieldCorrFuncAnalyl}).

One could use Fig.~\ref{Fig:gSpatialCorrelation} as a dark matter signature.  A natural question is how to probe various distances.
 In galactic coordinates, the Sun moves through the halo in the $(l=90^{\degree}, b=0^{\degree})$  direction, roughly towards the Cygnus constellation.
 The most straightforward approach is to place two satellites with, for example,
two precision clocks onboard in the vicinity of a Lagrange point, orient 
their relative position vector $\mathbf{d}$ towards Cygnus constellation (see Fig.~\ref{Fig:MapSatsEarth}(b)) and
carry out a series of measurements at various distances. This would map out
the spatial part of the correlation function. 

More practical approach is to carry out terrestrial experiments (see Fig.~\ref{Fig:MapSatsEarth}(c)) with fixed
positions of network nodes and rely on the angular dependence of the
$\mathbf{k}_{g}\cdot\mathbf{d}= \cos\theta_{g} \, m_{\phi}v_{g}d/\hbar $ of the phase. The ``galactic'' orientation angle $\theta_{g}$  would
change due to the Earth rotation. 
The inclination angle of the Cygnus constellation (i.e., the direction of motion through the  halo) as seen from the Earth
is about $45^\circ$. As an illustration, consider two geographically remote laboratories
located along the 45th parallel (see Fig.~\ref{Fig:MapSatsEarth}(c)). Their two-node correlation
function would sample  $\mathbf{k}_{g}\cdot\mathbf{d}$ in the range from
zero to  $k_{g}d/\sqrt{2}$ over the course of one day. It is worth
emphasizing that the coherence length $\lambda_{c}\sim k_{g}^{-1}$,
thereby two sites have the most sensitivity for $d<\lambda_{c}$.  

The geographic arrangement in Fig.~\ref{Fig:MapSatsEarth}(c) serves as an illustrative example as
simply time-synchronized measurements in existing laboratories~\cite{BudDer2015_PhysicsToday}, Fig.~\ref{Fig:MapSatsEarth}(a), should be
sufficient for the proposed global data analysis. GNOME network~\cite{Pustelny:2013rza} infrastructure can serve as a natural host for the VULF DM  observatory.
 One could also use clocks on
numerous navigational satellites, such as GPS, to search for correlation
patterns~\cite{DerPos14,Roberts2017-GPS-DM}; these have an advantage of half-a-day orbits (GPS) and a large
$ d \sim$50,000 km aperture.   eLISA gravitational wave  mission~\cite{Vitale2014}  (a network of three satellites) can also be used for VULF detection.
Another point is that the network does not need to be
homogenous, and various precision measurement tools can be included in the
network. Indeed, as long as the DM portal is linear, Eq.(\ref{Eq:LinearPortal}), such a global
observatory  can ``cast a much wider net'' on possible DM couplings. 

\section{Data analysis considerations}
\label{Sec:Constraints}
\subsection{ Statistical significance of the dark matter line shape}
\label{Sec:1DeviceStatistics}
Now I would like to establish  constraints on the coupling constants $\gamma_{X}$ of the linear DM-SM portal~(\ref{Eq:LinearPortal}).  
The temporal data stream $d_{k}=s_{k} + n_{k}$ is composed of the DM signal $s_{k}$ and the device noise $n_{k}$, $k \in [0,N-1]$.  I parameterize the DM signal as $s_{k} =\gamma_{X}  A \, \phi(t_{k})$, where the constant $A$ depends on the device (see an example for atomic clocks in Sec.~\ref{Sec:Clocks}), and $t_{k}=k \Delta_{t}$ refers to the time of the measurement $k$ with $\Delta_{t}$ being the sampling interval. For simplicity, I assume that the network is fixed in the halo reference frame; otherwise the positions of the terrestrial nodes would need to be tracked and the time-series transformed.
The noise quasi-PSD is defined as 
$\tilde\rho_{p} = \langle |\tilde{n_{p}}|^{2} \rangle$, where 
$\tilde{n}_{p}$ stands for discrete Fourier transformed (DFT) set of $n_{k}$, $\tilde{n}_{p} = \sum_{k=0}^{N-1} \exp( -i \frac{2 \pi}{N} kp) \, n_{k}$.
Review of DFT can be found in Appendix~\ref{Eq:LinearPortal}.
 The DM signal PSD can be expressed as 
$\langle |\tilde{s_{p}}|^{2} \rangle = A^{2} \gamma_{X}^{2} \langle |\tilde{\phi_{p}}|^{2} \rangle$ which can be further linked to the profile~(\ref{Eq:DMLineProfile}) via  
\begin{equation}
\langle |\tilde{\phi_{p}}|^{2} \rangle =  \frac{\pi N }{\Delta_{t}} \Phi_{0}^{2}  F\left(  \omega_{p} \right)\,, \label{Eq:DFTLineShape}
\end{equation}
 as long as there is no the DFT-inherent aliasing, see Sec.~\ref{Sec:Aliasing}.  Here $\omega_{p} = \frac{2\pi}{N \Delta_{t}} p$ is the DFT angular frequency.

The probability density for the field DFT coefficients  is given by the Rayleigh distribution (see Appendix~\ref{Sec:App:Stochastic})
\begin{equation}
 p( \tilde{\bm{\phi}} ) = \prod_{p=0}^{N/2} \frac{1}{ (\beta_{p}^{-1}\pi \langle |\tilde{\phi_{p}}|^{2} \rangle)^{\beta_{p}} } \exp\left( - \beta_{p} \frac{ |\tilde{\phi_{p}}|^{2} } {\langle |\tilde{\phi_{p}}|^{2} \rangle}  \right) \, .
 \label{Eq:FieldDistro}
 \end{equation}
In this expression, $\tilde{\bm{\phi}}$ is a vector composed of the field DFT components $\tilde{\phi}_{p}$, $\beta_{j} =1$ except for the DC and the Nyquist components for which 
$\beta_{0}=\beta_{N/2}=1/2$. 
A particular field realization can be constructed by randomly drawing coefficients $\tilde{\phi}_{p}$ from the distribution~(\ref{Eq:FieldDistro}) and 
using the inverse DFT, $\phi_{k} = N^{-1} \sum_{p=0}^{N-1}  \exp( +i \frac{2 \pi}{N} kp) \, \tilde{\phi}_{p}$.  The cause for the field decoherence is dephasing when multiple oscillations of different frequencies are added together.

The relevant likelihood for a stochastic signal is given by Eq.~(\ref{Eq:App:DFTstoch}). It is obtained by multiplying the DFT likelihood for a noisy device and a DM signal with fixed (deterministic) $\tilde{\bm{\phi}}$ with the field probability distribution~(\ref{Eq:FieldDistro}) and marginalizing over $\tilde{\bm{\phi}}$. The resulting stochastic signal likelihood can be recast into a  posterior probability density for the coupling strength $\gamma_{X}$. As shown in Appendix~\ref{Sec:App:CouplingStrength}, in the weak signal limit, the  maximum-likelihood estimator for $\gamma_{x}$ reads
\begin{equation}
 \hat{\gamma}_{X}^{(1)} = A^{-1} \left\{ 
 \frac{
\sum_{p=1}^{N/2-1}  \frac{\langle |\tilde{\phi_{p}}|^{2} \rangle}{ \tilde{\rho}_{p}} \left( \frac{ |\tilde{d}_{p}|^{2}}{\tilde \rho_{p} } -1  \right)
} {
 \sum_{p=1}^{N/2-1}  
 \left( \frac{\langle |\tilde{\phi_{p}}|^{2} \rangle}{ \tilde{\rho}_{p}} \right)^{2} \left( 2 \frac{ |\tilde{d}_{p}|^{2}}{\tilde \rho_{p} } -1  \right)
 }
\right\}^{1/2} \, ,
\label{Eq:gammaHat1}
\end{equation}
with the standard deviation 
\begin{equation}
 \hat{\sigma}^{(1)}_{\gamma_{X}} \approx A^{-1} \left\{  
 \sum_{p=1}^{N/2-1}  
 \left( \frac{\langle |\tilde{\phi_{p}}|^{2} \rangle}{ \tilde{\rho}_{p}} \right)^{2} \left( 2 \frac{ |\tilde{d}_{p}|^{2}}{\tilde \rho_{p} } -1  \right) 
 \right\}^{-1/4}\,.
 \label{Eq:sigmaHat1}
\end{equation}
 To streamline the notation, here I assumed that the DC ($p=0$) and the Nyquist ($p=N/2$) components  of the time series have been removed or filtered out. 
Further averaging~(\ref{Eq:sigmaHat1}) over multiple data realizations leads to an estimate
\begin{equation}
 \sigma^{(1)}_{\gamma_{X}} \approx A^{-1} \left\{  \sum_{p=1}^{N/2-1} \left( \frac{\langle |\tilde{\phi_{p}}|^{2} \rangle}{ \tilde{\rho}_{p}}\right)^{2} \right\}^{-1/4}\,.
 \label{Eq:sigma1}
\end{equation}
The constraint on $|\gamma_{X}|$  at the 68\% confidence level is $|\gamma_{X}| < \sigma^{(1)}_{\gamma_{X}}$.
An immediate consequence of Eq.~(\ref{Eq:sigma1}) is that the constraints on $|\gamma_{X}|$ scale with the number of frequency points $N_{F}$ sampled inside the DM line shape, Fig.~\ref{Fig:DMprofile},  as $1/N_{F}^{1/4}$. This is a qualitatively expected result, since the standard deviation of an average of $N_{F}$ data points (here the measured quantity is the VULF signal PSD), scales as $1/\sqrt{N_{F}}$. The additional square root comes from the fact that the VULF signal PSD $\propto (\gamma_{X})^{2}$. The number of points inside the profile $N_{F}$ is on the order of $(\tau_{c} \Delta_{f})^{-1}$, where  $\Delta_{f} = 1/(N\Delta_{t})$ is the DFT frequency step. 
Thus $N_{F} \sim N \Delta_{t}/\tau_{c}$.

The constraint~(\ref{Eq:sigma1}) can be simplified further for white noise devices, see Appendix~\ref{Sec:App:SumToIntegral} for details. For white noise, the PSD is flat, $\tilde \rho_{p} = N \sigma^{2}$, with $\sigma$ being the noise standard deviation.  For the fiducial value of $\eta=1$
\begin{equation}
|\gamma_{X}| < 2.4 \frac{\sigma}{A \Phi_{0}} \left(  \frac{\Delta_{t}}{N\tau_{c}} \right)^{1/4} \label{Eq:sigma1Simp}, 
\end{equation}
which formalizes the earlier qualitative observations. 
  The factor  $\left( \Delta_{t}/\tau_{c} \right)^{1/4}$ can be substantial: for Compton frequencies on the order of the sampling rate it is $O(\xi^{1/2} ) \approx 3 \times 10^{-2}$. 
For a fixed duration of a measurement campaign, the tightest bounds are obtained for $\Delta_{t} \sim \tau_{c}$.    
  The constraints can be further re-expressed in terms of boson masses and DM energy density,
\begin{equation}
|\gamma_{X}| < 5.4 \times 10^{-2} \frac{\sigma}{A } \frac{m_{\phi} c}{\hbar \sqrt{\rho_\mathrm{DM}}} \left( \frac{m_{\phi} c^{2}}{\hbar N} \Delta_{t} \right)^{1/4} \,,\label{Eq:sigma1Int}
\end{equation}
revealing the $m_{\phi}^{5/4}$ scaling when the device constant $A$ does not depend on the Compton frequency.

Now we consider a special case of a fully coherent signal, i.e., $\tau_{c} \gg N\Delta_{t}$. Then the field $\phi(t) = \Phi_{0} \cos( \omega_{\phi} t + \varphi)$, where $\varphi$ is some random but fixed phase.  For illustration, take  $\omega_{\phi}$ to be equal to one of the DFT angular frequencies, say $\omega_{m}$, $0<m<N/2$.
Then the  field PSD $\langle |\tilde{\phi_{p}}|^{2} \rangle =\frac{1}{4} \Phi_{0}^{2} N^{2} \delta_{p,m}$ and, for white noise, Eq.~(\ref{Eq:sigma1}) reduces to
\begin{equation}
|\gamma_{X}| < \sigma^{(1),\mathrm{coh}}_{\gamma_{X}} \approx \frac{\sigma}{A \Phi_{0}} \frac{2}{\sqrt{N}}.
 \label{Eq:sigma1coh}
\end{equation}
Thus comparing the stochastic field~(\ref{Eq:sigma1Simp}) and the coherent field~(\ref{Eq:sigma1coh}) constraints, we see that the sensitivity in the stochastic field case differ by a factor $(N \Delta_{t}/\tau_{c})^{1/4}$.  This is in  contradiction to the statement made in Ref.~\cite{ArvHuaTil15} (using notation of this paper)
``we expect to boost the sensitivity $\sigma$ of a single measurement by a factor of  $(\min(N, \tau_{c}/\Delta_{t}))^{1/2}$ -- the square root of the number of coherent measurements.''  
The correct statement would have included the factor of  
 $ (N \tau_{c} /\Delta_{t})^{1/4}$  instead of $(\tau_{c}/\Delta_{t})^{1/2}$   (see Eqs.~(\ref{Eq:sigma1Simp},\ref{Eq:sigma1coh}) for  additional $O(1)$ numerical factors).
 While the derived scalings are consistent with magnetometry sensitivity estimates~\cite{casper}, they apply to a broader range of experiments and include numerical factors specific to DM velocity distributions.

Based on Eqs.~(\ref{Eq:gammaHat1},\ref{Eq:sigmaHat1}), one could also devise signal-to-noise ratio (SNR) statistic in frequency space, 
\begin{equation}
\mathrm{SNR} \equiv  \frac{\hat{\gamma}_{X}^{(1)}}{\hat{ \sigma}^{(1)}_{\gamma_{X}}} \, ,
\label{SNR}
\end{equation}
that can be applied to the data streams directly. This is an application of the  matched filter technique (for example, used by the gravitational wave community~\cite{RomanoCornish2017}). The specific values of the VULF PSD $\langle |\tilde{\phi_{p}}|^{2} \rangle$ template depend on the field Compton frequency. Thus the SNR statistic needs to be scanned against multiple templates by varying $f_{\phi}$ and searching for $\mathrm{SNR}$ values to exceed a certain value, e.g., $\mathrm{SNR}> 5$.

One of the subtleties is that, while deriving formulas in this section, the device noise PSD $\tilde\rho_{p}$ was assumed to be known. The VULFs are, however, ever-present and can not be  turned off or shielded out. A strategy could be to fit the data PSD with a smooth polynomial, and assign the noise PSD to the smooth background. In addition, more sophisticated, Bayesian logic-based approaches can be adopted, see, e.g., mixture model discussion in Sec.~23 of Ref.~\cite{Linden-BayesianBook2014}.  

Finally, it is worth noting that  Eqs.~(\ref{Eq:sigma1Simp},\ref{Eq:sigma1Int}) were derived in the assumption that 
the entire DM line shape is sampled in the DFT frequency space, i.e., nominally for VULF frequencies below the DFT Nyquist frequency. 
However, as shown in the following section, this limitation can be  substantially relaxed and the sensitivity to Compton frequencies above the Nyquist frequency can be gained due to {\em aliasing} inherent to the DFT. I will demonstrate that the sensitivity estimates (\ref{Eq:sigma1Simp},\ref{Eq:sigma1Int}) hold for  $f_{\phi} \lesssim 10^{6}/(2 \pi \Delta_{t})$.

\subsection{Probing high-frequency dark matter fields through aliasing}
\label{Sec:Aliasing}
A high-frequency (above the Nyquist frequency) DM field still interacts with the device, but it is just sampled at an insufficient rate to resolve individual  field oscillations. The aliasing in discrete sampling qualitatively refers to the fact that the PSD of an oscillation of frequency $\omega_{\phi}$ and that of same amplitude and phase  oscillations  of frequencies $\omega_{\phi}+(2\pi/\Delta_{t}) q$, with $q$ being an integer, are identical. In other words, the DM oscillations with frequencies above the Nyquist frequency are effectively shifted to the nominal DFT range: $f_{\phi,\mathrm{aliased}} = \mod(f_{\phi},1/\Delta_{t})$.  To explicitly demonstrate this relation,
examine the statement of the Wiener-Khinchin theorem in DFT space relating the PSD and the time-domain correlation function
\begin{equation}
\langle |\tilde{\phi_{p}}|^{2} \rangle = N \sum_{k=-(N-1)}^{N-1} \exp( - i \frac{ 2\pi}{N} p k ) g\left( k \Delta_{t}  ,\mathbf{0}\right).
\label{Eq:WKT-DFT}
\end{equation}
It can be verified by a direct substitution in Eq.~(\ref{Eq:LocalCorrelation}) that $g\left( k \Delta_{t} ,\mathbf{0}\right)$ is invariant under $\omega'_{\phi} \rightarrow \omega'_{\phi}+(2\pi/\Delta_{t}) q$. 

\begin{figure}[!ht]
\begin{center}
\includegraphics[width=0.9\columnwidth]{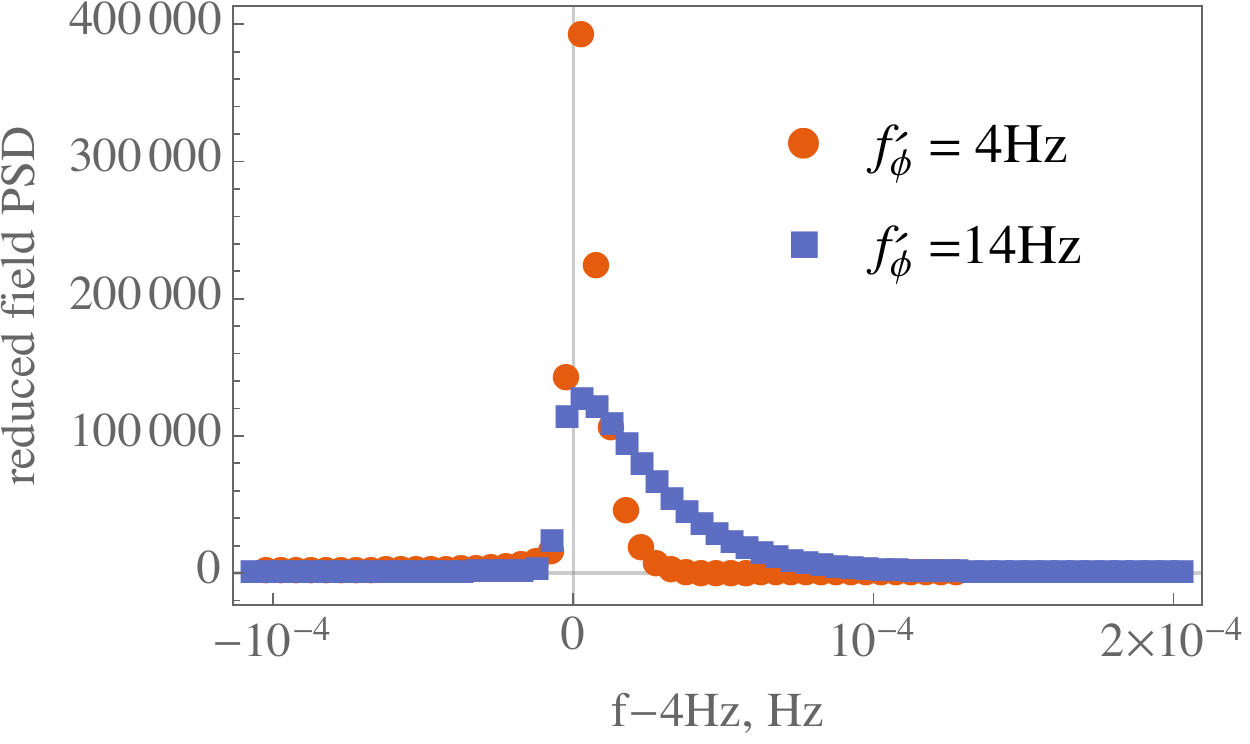}
\includegraphics[width=0.9\columnwidth]{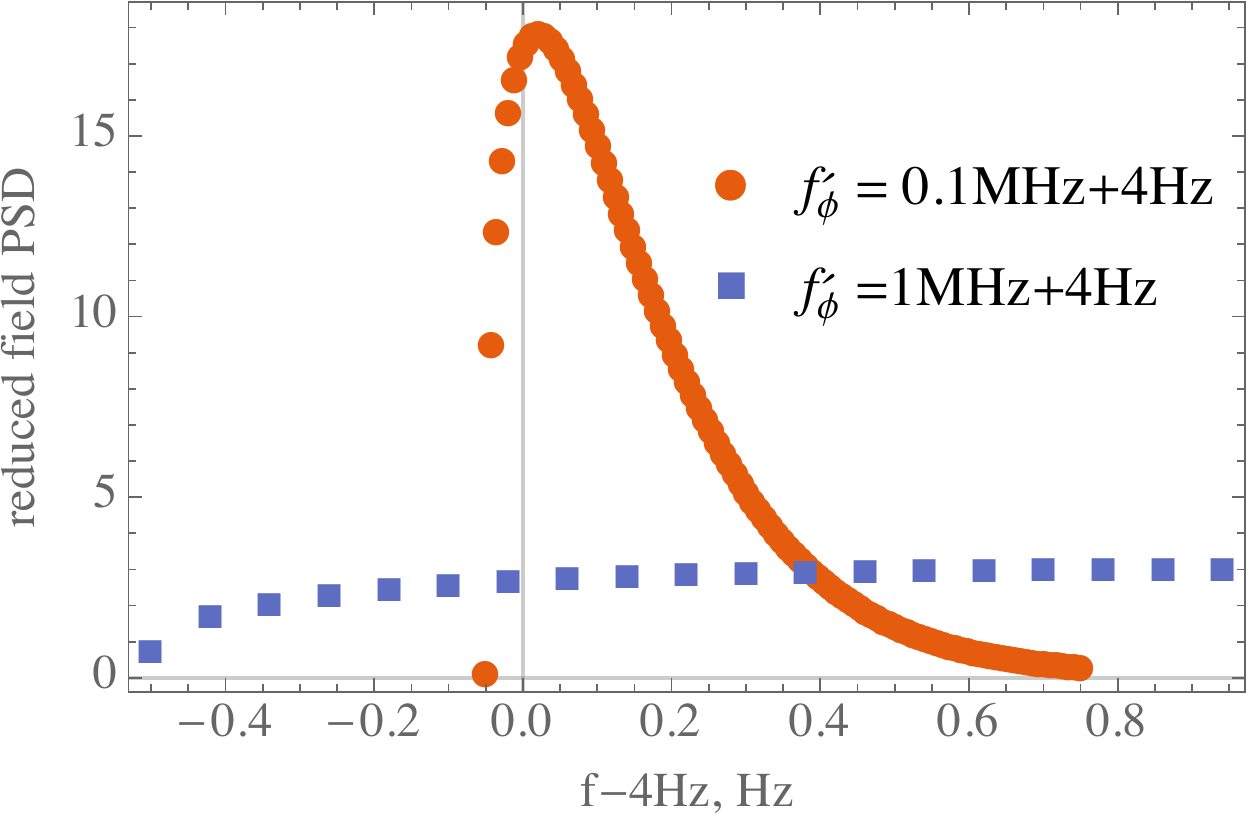}
\caption{ \label{Fig:aliasing} Effects of DFT aliasing on the observable  power-spectral density of dark matter field. Reduced PSD is
defined as $\langle |\tilde{\phi_{p}}|^{2} \rangle/(N \Phi_{0}^{2})$ and is shown as a function of DFT frequency detuning from 4 Hz. 4 Hz is the aliased Doppler-shifted Compton frequency for all curves. Nyquist frequency in the simulation is 5 Hz. See text for other simulation parameters. The top panel is for low-frequency Doppler-shifted Compton frequencies and the bottom panel shows PSDs for high-frequency fields. The data sampling rate is the same for all shown curves.
}
\end{center}
\end{figure}

Fig.~\ref{Fig:aliasing} demonstrates the aliasing effect on the sampled DM field PSD for various Compton frequencies.
In this figure, to generate the field PSDs, I carried out summations in the  Wiener-Khinchin theorem~(\ref{Eq:WKT-DFT}) numerically with 2-point correlation function~(\ref{Eq:LocalCorrelation}) setting $v_{g}=v_\mathrm{vir}$.  The coherence time $\tau_{c}$ was  varied with Compton frequency according to Eq.~(\ref{Eq:PracticalTauC}).
In this simulation, the total number of points  and the sampling interval were  $N=2\times10^{6}$ and  $\Delta_{t}=0.1\, \mathrm{s}$.  The corresponding Nyquist frequency is $f_\mathrm{Nyquist}=1/(2\Delta_{t}) = 5\, \mathrm{Hz}$.  Several observations can be made based on this simulation:
\begin{itemize}
\item[(i)] While the high-frequency fields are aliased to the nominal DFT frequency range, the original coherence time is retained by the aliased copy. {\em Therefore, an experimentalist should search for aliased VULF lines 
that may be broader than the ones expected from the VULF oscillations in the nominal $0 \leq f_{\phi} \leq f_\mathrm{Nyquist}$ frequency range.} 
\item[(ii)] The shape of the DM line is preserved as long as the bulk of the aliased profile fits inside  the nominal DFT frequency range $0\leq f \leq f_\mathrm{Nyquist}$. If $\tau_{c} \lesssim \Delta_{t}$, the shape is distorted by the aliasing, and the background-signal separation becomes challenging. An example of such a distortion is for $f'_{\phi} = 1 \,\mathrm{MHz} + 4 \mathrm{Hz}$ curve in Fig.~\ref{Fig:aliasing}. This observation, in particular, means that the sensitivity estimates (\ref{Eq:sigma1Simp},\ref{Eq:sigma1Int}) hold for  $f_{\phi} \lesssim 10^{6}/(2 \pi \Delta_{t})$.
\item[(iii)] The ``non-distortion'' condition $\tau_{c} \lesssim \Delta_{t}$ means that the VULF field must remain coherent over a duration of a sampling period. 
\end{itemize} 

Suppose the total observation time $T=N\Delta_{t}$ is smaller than the coherence time for Compton frequency that is below $f_\mathrm{Nyquist}$. Then this oscillation behaves as a deterministic coherent signal on the time scale $T$. As we start searching for
higher Compton frequencies via aliasing, the DM signal becomes increasingly stochastic in nature, with coherence time on the order of the sampling time $\Delta_{t}$ for the highest probed frequencies. Then the stochastic field approach developed in this paper, while being valid for deterministic signals, becomes especially relevant.  Statistical estimators~(\ref{Eq:gammaHat1},\ref{Eq:sigmaHat1},\ref{Eq:sigma1}) explicitly refer to the DM field PSD $\langle |\tilde{\phi_{p}}|^{2} \rangle$. In practice, these quantities should be estimated using DFT sampling~(\ref{Eq:WKT-DFT}) of the correlation function  to properly reflect aliasing and the accompanying distortions. Notice that the use of Doppler shifted frequencies for high-frequency fields is important, because aliased frequencies are mapped as $\mod(f'_{\phi},1/\Delta_{t})$, and $1/\Delta_{t}$ may be comparable to the Doppler shift $m_{\phi} v^{2}_{g}/(4 \pi \hbar)$.
 
Aliasing effects extend the DM searches to higher Compton frequencies, as long as the frequencies of VULF oscillations are within the device bandwidth.  This idea has been  previously employed in Ref.~\cite{Arvanitaki2016-GWdetectors} to establish projected high-frequency constraints on VULF coupling strengths in atom interferometry.

\subsection{Statistical significance of correlation function for a network}
\label{Sec:NetworkStatistics}
Sec.~\ref{Sec:1DeviceStatistics} established  constraints on the coupling strength $\gamma_{X}$ for a single device.
The derivation for a multi-node network is similar, starting from the network likelihood~(\ref{Eq:App:Network-Stochastic}), see Appendix~\ref{Sec:App:Network} for details. I  label the devices with the letters at the beginning of the alphabet $a,b,c,\ldots$ and operate with $N_{d}$-dimensional  quasi-PSD matrices $\bm{\rho}_{p}$ and $\bm{S}_{p}$, where index $p$, as previously, refers to the DFT frequency. 
The noise and  the DM signal  PSD matrices $\bm{\rho}_{p}$ and  $\bm{S}_{p}$ have  elements in the node space $\rho^{a,b}_{p}=\langle \tilde n^{a}_{p} \, (\tilde n^{b}_{p})^{*} \rangle$ and $S^{a,b}_{p}=\gamma_{X}^{2} A^{2} \langle \tilde \phi^{a}_{p} \,  (\tilde \phi^{b}_{p})^{*} \rangle$, respectively. $\tilde n^{a}_{p}$ is the DFT component of the $a$-th device noise   and $\tilde{\phi}^{a}_{p}$ is the DFT component of the field at the location $\mathbf{r}^{a}$ of the $a$-th device. If the devices are independent,  i.e., their noise is mutually uncorrelated  and the  devices are identical, $\rho^{a,b}_{p} = \delta_{a,b} \tilde{\rho}_{p}$. The quantities  $\varPhi^{a,b}_{p}\equiv \langle \tilde \phi^{a}_{p} \,  (\tilde \phi^{b}_{p})^{*} \rangle$ are related through the Fourier transform (the Wiener-Khinchin theorem)  to the derived two-point correlation function 
$g\left(  \tau,\mathbf{d}^{a,b}=\mathbf{r}^{a} -\mathbf{r}^{b} \right)$ .

The resulting standard deviation for the coupling strength $\gamma_{X}$ in the weak DM signal limit reads
\begin{equation}
\sigma_{\gamma_{X}}^{\mathrm{(network)}} = A^{-1} \left\{ 
\sum_{p=1}^{N/2-1} \frac{\mathrm{tr}({ \bm{\varPhi}_{p} \bm{\varPhi}}_{p}) }{ (\tilde{\rho}_{p})^{2}} 
\right\}^{-1/4} \,.
\label{Eq:sigmaNetwork}
\end{equation}
The major difference with the single device formula~(\ref{Eq:sigma1}) is the presence of the trace $\mathrm{tr}({ \bm{\varPhi}_{p} \bm{\varPhi}}_{p}) = 
\sum_{a,b} \varPhi^{a,b}_{p} \varPhi^{b,a}_{p}$. If all $N_{d}$ nodes are separated by distances larger than the coherence length $\lambda_{c}$, only  the diagonal elements contribute to the trace and $\sigma_{\gamma_{X}}^{\mathrm{(network)}} =  \sigma^{(1)}_{\gamma_{X}}/ N_{d}^{1/4}$, where 
$\sigma^{(1)}_{\gamma_{X}}$ is the standard deviation for a single device~(\ref{Eq:sigma1}). 

In the opposite limit of the node separations being much smaller than $\lambda_{c}$,
$\sigma_{\gamma_{X}}^{\mathrm{(network)}} =  \sigma^{(1)}_{\gamma_{X}}/ \sqrt{N_{d}}$. Thus, compared to an individual device, the statistical sensitivity of a fully coherent network is improved by the factor $\sqrt{N_d}$, where $N_\mathrm{d}$ is the number of nodes.
The qualitative reason for this scaling is that at each DFT frequency, the network samples $(N_{d})^{2}$ DM quantities $S^{a,b}_{p}$. Due to the conventional $1/\sqrt{\textrm{(number of data points})}$ averaging dependence, this leads to $\sigma_{\gamma_{X}}^{\mathrm{(network)}} \propto 1/\sqrt{N_d}$.  An additional square root comes  from the fact that each of the measured quantities $S^{a,b}_{p}$ is proportional to  $(\gamma_{X})^{2}$. 

The  best bounds are attained for co-located devices or the fully coherent network, i.e., when $d^{a,b} \ll \lambda_{c}$.
Indeed, in this case in Eq.~(\ref{Eq:sigmaNetwork}) all cross-node correlators are equal to a single node correlator: $\varPhi^{a,b}_{p} =  \varPhi^{a,a}_{p}$. For a distributed network, however, $\varPhi^{a,b}_{p} \leq \varPhi^{a,a}_{p}$ and the statistical sensitivity is reduced
and is bounded by the case when $d^{a,b} \gg \lambda_{c}$ for all the network links:
$$
\sigma^{(1)}_{\gamma_{X}}/ (N_{d})^{1/2} \le \sigma_{\gamma_{X}}^{\mathrm{(network)}} \le \sigma^{(1)}_{\gamma_{X}}/ (N_{d})^{1/4} 
\,.
$$

An experiment operating at a Hz sampling rate probes sub-Hz Compton frequencies (see, however, Sec.~\ref{Sec:Aliasing} for the possibility of probing higher frequencies), which translate into coherence lengths, Eq.(\ref{Eq:CohLengthPractical}), $\lambda_{c} \gtrsim 5 \times 10^{7} \, \mathrm{km}$. Thus even a global terrestrial network ($d^{a,b} \lesssim 10^{4} \, \mathrm{km}$)  of such devices would operate in the ``fully coherent'' regime, gaining the most in sensitivity from the correlated analysis: $\sigma_{\gamma_{X}}^{\mathrm{(network)}} =  \sigma^{(1)}_{\gamma_{X}}/ \sqrt{N_{d}}$. This condition, however, can change
for higher-frequency fields  with shorter coherence times probed through the aliasing technique. For example, for a MHz field, the coherence length is $\sim 10\, \mathrm{km}$.



\section{Atomic clocks}
\label{Sec:Clocks}
In this section, I qualitatively discuss an application of the presented formalism  to atomic clocks and their networks.

In our preceding discussion we assumed that the measurements were instantaneous; in practice, there is always a finite interrogation time $t_0$ for a single measurement. I assume that the next measurement is taken right after the previous 
one was completed (no ``dead'' times), with DFT sampling time interval of Sec.~\ref{Sec:Constraints} $\Delta_{t} =t_{0}$. We form a time series of fractional frequency excursions  
$s^{(a)}_n \equiv (\omega_n-\omega_c)/\omega_c$ taken at $t_n=n t_0; n=\overbar{1,N}$ for a fixed inter-node distance $d$, with $a$ labeling the node and $\omega_c$ being the nominal clock frequency. The VULF contribution to 
$s^{(a)}_n$ can be expressed in terms of the single-measurement accumulated clock phase
and sensitivity coefficients $K_X=\partial \ln \omega_c/\partial \ln X -\partial \ln \omega_\mathrm{LO}/\partial \ln X $, where $\omega_\mathrm{LO}$ is the resonance frequency of the local oscillator (reference cavity):
\begin{equation}
s^{(a)}_n =  \left( \sqrt{\hbar c} \sum_X{ \gamma_{X} K_X } \right)  \int_{t_{n}-t_0}^{t_n} \phi(\mathbf{r}^{a},t') \frac{dt'}{t_0}  . 
\label{Eq:ClockSignal}
\end{equation}
Notice the integral of the VULF field time evolution history over the interrogation duration. In this section, $X$ runs over fundamental constants that affect the atomic or the local oscillator resonance frequencies. The constants may include the electromagnetic fine-structure constant $\alpha$, mass of the electron $m_{e}$ and so on,  see, e.g., discussion~\cite{Roberts2017-GPS-DM} for both optical and microwave clocks. To streamline the notation, we combine $\sum_X{ \gamma_{X} K_X } =\gamma_\mathrm{eff}$, and $\gamma_{X}$ is to be replaced with  $\gamma_\mathrm{eff}$ in the 
previous sections.  

The reference to the dependence of the local cavity resonance frequency $\omega_\mathrm{LO}$ on fundamental constants in the sensitivity coefficients $K_{X}$ is due to the fact that 
in the Ramsey interrogation scheme, the accumulated atomic  phase and thus the detected quantum probability of a resonant transition is determined by a time integral of the difference in frequencies between the clock atom and the local oscillator. The local oscillator (reference cavity) itself is a subject to the DM field influence during the interrogation. An example of such an effect is the DM-induced variation in the Bohr radius $a_{0} = \alpha \hbar/(m_{e}c) $ affecting cavity length $L \propto a_{0}$ and thus the cavity resonance frequencies~\cite{StaFla2015}.  I refer the reader to further clock-specific discussions~\cite{Wcislo2016,Roberts2017-GPS-DM}.
 
Eq.~(\ref{Eq:ClockSignal}) is an approximation as it assumes instantaneous responses of the atom and the local oscillator to the time-varying fundamental constants. In particular, the cavity response is not instantaneous, for example, due to the laser pulse  intra-cavity round-trip time $2L/c$ being finite. Moreover, the macroscopic adjustment of the macroscopic cavity length to the microscopic variations of fundamental constants  requires propagation of sound waves in the cavity spacer material~\cite{Arvanitaki2016_SoundDM}, and, therefore, the cavity dynamic response can exhibit a  frequency cutoff above the characteristic frequency $v_{s}/L\sim 10^{5} \, \mathrm{Hz}$, where $v_{s}$ is the speed of  sound in the cavity spacer. Dynamic  response of an atom is much faster, with an expected qualitative change in the otherwise nearly instantaneous response behavior above typical atomic frequencies, which are comparable to the upper limit on the plausible values of Compton frequencies,
see Table~\ref{Tab:VULF-params}. Thereby, in general, the sensitivity coefficients $K_{X}$ also depend on time and, strictly speaking, should include the delayed cavity response. Another subtlety is an operation of the atomic clock servo-loop that locks the cavity resonance to the atomic frequency and requires several measurement cycles, see Ref.~\cite{Roberts2017-GPS-DM}.

It is worth mentioning that it is  hardly necessary to make frequency ratio comparisons between clocks of different sensitivities to the variation of fundamental constants as in the original proposal~\cite{ArvHuaTil15} and the follow-up experimental work~\cite{VanTilburg2015,Hees2016}. The frequency comparison of a local oscillator and the clock atoms is naturally carried out per the conventional operation of a single atomic clock and it is sufficient. Another advantage of using a single clock is a larger sensitivity to the variation of fundamental constants (see also Ref.~\cite{Arvanitaki2016-GWdetectors}). For both microwave and optical clocks $\partial \ln \omega_c/\partial \ln \alpha \approx 2$ due to the Rydberg constant being $\alpha^{2}m_{e}c^{2}$.  In the frequency-ratio technique, however,
this dominant contribution cancels out and the sensitivity is  attributed entirely to the small difference in $\alpha$-dependent relativistic corrections to atomic structure.

Now, assuming the validity of Eq.~(\ref{Eq:ClockSignal}),  consider a measurement with a single clock. If the  DM field remains coherent over a duration of a single measurement, $t_{0} \ll \tau_{c}$, the field behaves as 
$\phi(\mathbf{r}^{a},t) \approx \Phi_{0} \cos(  \omega'_{\phi} t + \varphi^{a})$ over $t_{0}$, 
where the phase $\varphi^{a} = \varphi - \mathbf{k} \cdot \mathbf{r}^{a}$, with 
$\varphi$ being a fixed phase common to all nodes. Then the VULF signal, Eq.~(\ref{Eq:ClockSignal}), is explicitly
\begin{equation}
s^{(a)}_n =  \sqrt{\hbar c} \gamma_\mathrm{eff}  \frac{\sin( \omega'_{\phi} t_{0}/2)}{\omega_{\phi} t_{0}/2} \phi(\mathbf{r}^{a},t_{n-1} + t_{0}/2), \label{Eq:clockSignal}
\end{equation}
leading to the identification of the device constant $A$ of Sec.~\ref{Sec:Constraints} as $A =  \sqrt{\hbar c} \, \mathcal{W}( \omega'_{\phi} t_{0})$, where  the ``filter function''   is
\begin{equation}
\mathcal{W}( \omega'_{\phi} t_{0}) = \frac{\sin( \omega'_{\phi} t_{0}/2)}{\omega'_{\phi} t_{0}/2}.
\label{Eq:Filter}
\end{equation}
This function emphasizes the dependence
on the ratio of the interrogation time to the period of VULF oscillation. If VULF oscillations are slow compared to $t_0$, then  $\mathcal{W}( \omega_\phi' t_0 \ll 1) \approx 1$ and if they are fast, the effect tends to average out,  $|\mathcal{W}( \omega_\phi' t_0 \gg 1)| < 2 (\omega_\phi' t_0)^{-1}$. For a typical~\cite{DerKat11} $t_0 \sim 1 \,\mathrm{s}$ for optical lattice clocks,  the separation between the two regimes occurs at $f_\phi \approx 1 \, \mathrm{Hz}$ ($m_\phi \sim 10^{-14} \, \mathrm{eV}$). For the nominally probed Compton frequencies $f_{\phi} < 1/(2 t_{0})$, the influence of the window function is minimal, $\mathcal{W}( \omega_\phi' t_0) \approx 1$.
However, the bandwidth-limiting effect of the filter function becomes important for higher frequencies  probed through the aliasing technique,
Sec.~\ref{Sec:Aliasing}. The condition $t_{0} \ll \tau_{c}$ employed in deriving Eq.~(\ref{Eq:clockSignal}) is consistent with the limits on applicability of the aliasing technique, $f_{\phi} \lesssim 10^{6}/(2 \pi t_{0})$. For our illustrative optical lattice clock example with $t_{0}=1\,\mathrm{s}$, $f_{\phi} \lesssim 0.2 \, \mathrm{MHz}$ ($m_{\phi} \lesssim 8 \times 10^{-10}\,\mathrm{eV}$).

The search in the indicated VULF frequency parameter space can be carried out by directly employing the frequency-space SNR statistic of Sec.~\ref{Sec:1DeviceStatistics}. If the SNR $<1$ for the entire probed frequency space, one could rigorously constraint the coupling constant $\gamma_\mathrm{eff}$ through Eq.~(\ref{Eq:sigma1}) for the measured clock noise PSD.

 An estimate of the sensitivity can be made assuming that the   
measurement noise is dominated by the white frequency noise, i.e.,  Allan variance scales as $\sigma_{y}(\tau) \propto 1/\sqrt{\tau}$.  The associated noise PSD is flat. Then $\sigma$ in Eq.~(\ref{Eq:sigma1Int})  can be replaced by the Allan variance, $\sigma_{y} (t_{0})$, leading to the constraint on the effective coupling strength
\begin{equation}
|\gamma_\mathrm{eff}| < 5.4 \times 10^{-2} \frac{\sigma_{y}(t_{0})}{|\mathcal{W}( \omega_{\phi} t_{0})|} \frac{m_{\phi} c^{1/2}}{\hbar^{3/2} \sqrt{\rho_\mathrm{DM}}} \left( \frac{m_{\phi} c^{2}}{\hbar N} t_{0} \right)^{1/4} \,.\label{Eq:sigma1ClockFinal}
\end{equation} 
This estimate holds for $t_{0} \ll \tau_{c}$ and $\tau_{c} \ll N t_{0}$. For low frequencies, $\tau_{c} \gg N t_{0}$,  one needs to employ properly adopted Eq.~(\ref{Eq:sigma1coh}).

\begin{figure}[!ht]
\begin{center}
\includegraphics[width=1.0\columnwidth]{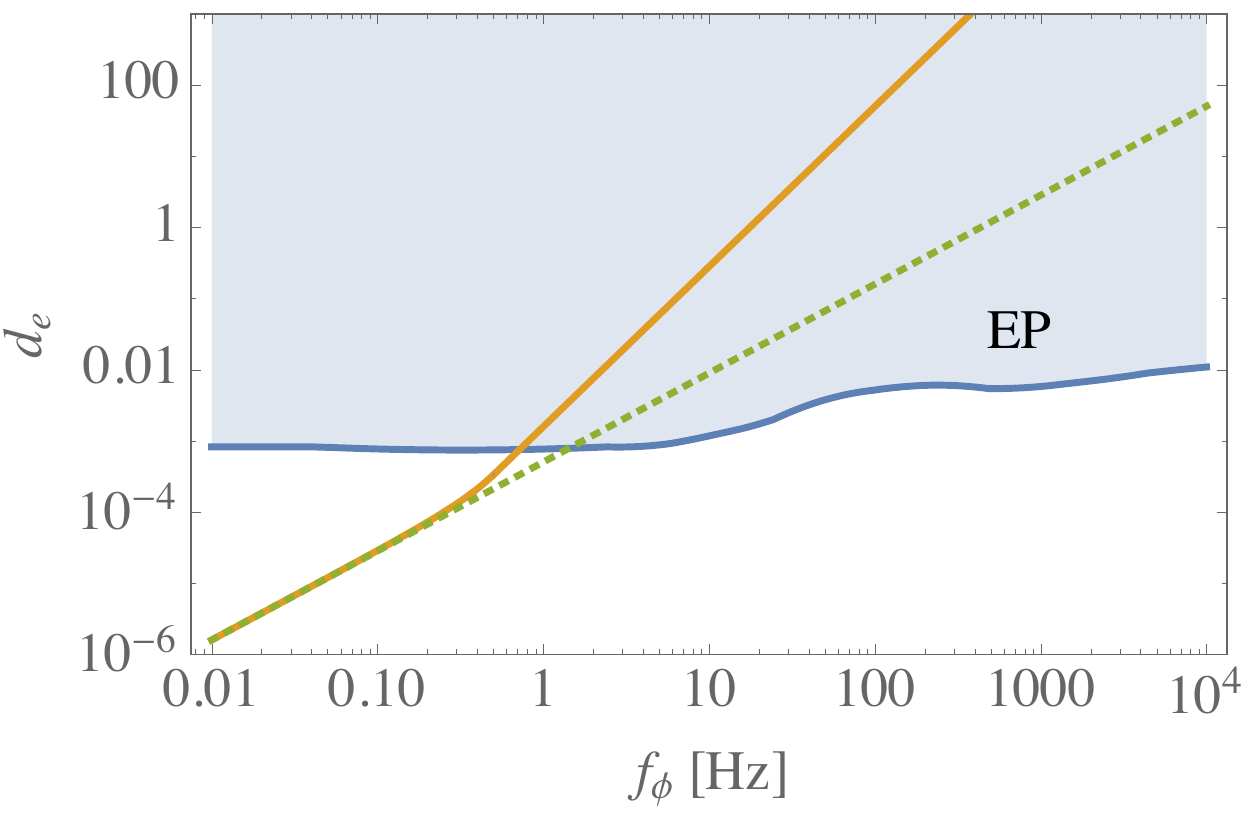}
\caption{ \label{Fig:limits} 
Projected constraints on the electromagnetic gauge modulus  $d_{e}$ for high Compton frequencies. The excluded parameter space from the Equivalence  Principle (EP) tests~\cite{Arvanitaki2016_SoundDM} is shown as a shaded region. Solid orange line is the projected limit from Eq.~(\ref{Eq:sigma1ClockFinal}) for an optical clock compared  to the state-of-the-art cavity~\cite{Zhang2017} (see text for details). The dashed green line is drawn neglecting the bandwidth-limiting effect of the filter function~(\ref{Eq:Filter}).
}
\end{center}
\end{figure}

The projected constraints on  the electromagnetic gauge modulus  $d_{e}$ is shown in Fig.~\ref{Fig:limits}. I used the  value of $\sigma_{y} (1\, \mathrm{s}) = 10^{-16}$ characteristic of modern optical cavities~\cite{Zhang2017}. I  set $K_{\alpha} =3$ and  suppressed sensitivity to the electron mass variation, $d_{m_{e}}\equiv0$. Then  $\gamma_\mathrm{eff} = 3 \gamma_{\alpha} = 3 \sqrt{2 \pi} d_{e}/E_\mathrm{P}$, where the last relation is for the electromagnetic gauge modulus $d_{e}$, see Sec.~\ref{Sec:Portals}. For high  ($f'_{\phi} > 1/t_{0}$) frequencies I set $|\mathcal{W}( \omega_\phi' t_0)| = 2 (\omega_\phi' t_0)^{-1}$ as one could simply repeat the experiment with a slightly adjusted $t_{0}$ to maximize sensitivity for a given $\omega'_{\phi}$. The single measurement time $t_{0}$ was set to 1 s and one year of integration time was assumed. In Fig.~\ref{Fig:limits}, there is a region of parameter space, where the clocks can explore yet unconstrained parameter space. The existing equivalence principle constraints on the modulus  $d_{m_{e}}$ are about an order of magnitude worse than those for $d_{e}$~\cite{Arvanitaki2016_SoundDM}, somewhat improving the discovery reach of atomic clocks.  
To improve the sensitivity, one needs to improve the short-term clock stability.  While in the region of high frequencies, resonant mass detectors may be more competitive~\cite{Arvanitaki2016_SoundDM}, but are narrow-band.

The constraints can be improved further by employing a network of clocks. 
The  global clock networks include the Global Positioning System and other navigational satellite constellations and a trans-European network of laboratory clocks~\cite{Lisdat2015_clockNetwork,Riehle2017}.  Navigational satellites house microwave clocks on board and the navigational systems also include clocks on terrestrial stations. Several local clock networks are available at national metrology institutes around the world, mostly used for clock comparisons.
It is clear from the  discussion in Sec.~\ref{Sec:NetworkStatistics} that compared to an individual clock, the statistical sensitivity of a network is improved by a factor of up to $\sqrt{N_{d}}$, where $N_{d}$ is the number of clocks. The best limits on $\gamma_{X}$ are attained when the clocks are co-located (or within the field coherence length), but, in the event of a positive observation, the confidence is  improved for a distributed network.

\section{Conclusions}
\label{Sec:Conclusions}

Examination of the VULF parameter space in Table~\ref{Tab:VULF-params} raises a question: when can the DM signal  be considered deterministic and when it is stochastic in nature. If the total measurement time is below coherence time, then the VULF behaves as a deterministic oscillating signal. While the phase, the amplitude, and the frequency of the oscillations are randomly drawn from the probability distribution~(\ref{Eq:FieldDistro}), their values persist over the entire measurement duration. If, however, the duration of the measurement campaign is much longer than the VULF coherence time, then multiple field realizations are sampled and one needs to apply stochastic field techniques discussed in this paper.  The stochastic approach becomes even more valuable when the high-frequency fields are probed through the aliasing effect inherent to discrete sampling, as the coherence time becomes shorter for such fields. The presented formalism is applicable to both cases and can serve as a starting point for statistical analysis.

This paper established  theoretical formalism for analyzing  stochastic properties of ultralight dark-matter fields through an explicit evaluation of  DM field $N$-point correlation function. The correlation function encodes all so far established DM priors. 
For a single device, I derived a DM line-shape profile than can be directly compared to the experimental data through the  developed SNR statistic. The paper also addressed DM sensitivity of a distributed (or co-located) network of precision measurement tools.

\begin{acknowledgments}
I would like to thank A. Arvanitaki, D. Budker,  A. Geraci, P. Graham,  R. Guyer, D. Kimball, M. Lukin, M. Pospelov,  B. Roberts, and J. Weinstein for discussions and N. Rodrigue for help with the figures.
I would like to acknowledge helpful conversations with K. Smith about statistical analysis in frequency space during my visit at the Perimeter Institute. 
I am also grateful to the Mainz Institute for Theoretical Physics (MITP) for  hospitality.
This work was supported in part by the U.S. National Science Foundation, by the University of Nevada, Reno mICRo grant,
and by Perimeter Institute for Theoretical Physics. Research at Perimeter Institute is supported by the
Government of Canada through the Department of Innovation, Science and Economic Development and by the Province of Ontario through
the Ministry of Research and Innovation.
\end{acknowledgments}

\appendix

\section{Discrete Fourier transform and frequency-space probability distributions for deterministic and stochastic signals}
\label{Sec:App:DFT-probabilities}

\subsection{Review of Discrete Fourier Transform (DFT)}
\label{Sec:App:DFT}

Introductions to DFT can be found, for example in
Refs.~\cite{Gre10book,RomanoCornish2017}. Here we review key definitions and
relevant results, and also introduce notation used through the main text of
the paper. The author acknowledges helpful discussions with Kendrick Smith and
Joseph Romano.

Consider a time series $\left\{  x_{k}\right\}  $ with values tabulated at
times $t_{k}=k\Delta_{t},~k=\overline{0,N-1}$. The total observation time is
$T=N\Delta_{t}$, where $N$ is the total number of individual measurements and
$\Delta_{t}$ is the time step. We will assume that $N$ is an even number and
$x_{k}$ are real-valued. The discretized time series is continued periodically
outside the observation time interval.

The DFT components $\tilde{x}_{p}$ are defined via the discretized Fouried
transformation of the time series,
\begin{equation}
\tilde{x}_{p}=\sum_{k=0}^{N-1}\exp\left(  -i\frac{2\pi}{N}pk\right)
\,x_{k}\,.\label{Eq:DFT}%
\end{equation}
The index $p$ refers to the DFT frequencies $f_{p}=p/\left(  N\Delta
_{t}\right)  $ or, equivalently, to the DFT angular frequencies $\omega
_{p}=2\pi f_{p}$. The frequency steps are $\Delta_{f}\equiv1/\left(
N\Delta_{t}\right)  $ and $\Delta_{\omega}\equiv2\pi/\left(  N\Delta
_{t}\right)  $. 

Introducing vectors $\mathbf{x=}\left\{  x_{0},x_{1},\ldots,x_{N-1}\right\}
^{T}$ and $\mathbf{\tilde{x}=}\left\{  \tilde{x}_{0},\tilde{x}_{1}%
,\ldots,\tilde{x}_{N-1}\right\}  ^{T}$, DFT can be interpreted as a linear
transformation (basis rotation)
\begin{equation}
\boldsymbol{\tilde{x}}=\sqrt{N}\boldsymbol{U}~\boldsymbol{x}%
\,,\label{Eq:App:DFT-matrix}%
\end{equation}
where the elements of the matrix $\boldsymbol{U}$ are $U_{pk}=\frac{1}%
{\sqrt{N}}\exp\left(  -i\frac{2\pi}{N}pk\right)  $. The transformation matrix
$\boldsymbol{U}$ is symmetric, $\boldsymbol{U}^{T}=\boldsymbol{U}$, and
unitary, $\boldsymbol{U}^{-1}=\boldsymbol{U}^{\dagger}$. Using unitarity,
Eq.~(\ref{Eq:App:DFT-matrix}) can be inverted to yield the inverse DFT,
\begin{equation}
\boldsymbol{x}=\left(  \sqrt{N}\right)  ^{-1}\boldsymbol{U}^{\dagger
}~\boldsymbol{\tilde{x}}\,.\label{Eq:App:invDFT-matrix}%
\end{equation}

DFT values are periodic, $\tilde{x}_{N+p}=\tilde{x}_{p}$, as can be verified
directly from Eq.~(\ref{Eq:DFT}). Thereby, we limit $p$ to the
\textquotedblleft nominal\textquotedblright\ range $\overline{0,N-1}$ with
frequencies $f_{p}$ ranging from $0$ to $(N-1)/(N\Delta_{t})$. Since
$\mathbf{x}$ are real-valued,
\begin{equation}
\tilde{x}_{p}=\tilde{x}_{-p}^{\ast}=\tilde{x}_{N-p}^{\ast}%
\,.\label{Eq:DFTperiodicity}%
\end{equation}
Clearly, the information contained in coefficients $\tilde{x}_{p}$ at
frequencies above the Nyquist frequency, $f_{\mathrm{Nyquist}}=f_{N/2}%
=1/\left(  2\Delta_{t}\right)  $, is redundant as these coefficients can be
recovered by complex conjugating coefficients of frequencies below
$f_{\mathrm{Nyquist}}$. Another consequence of Eq.~(\ref{Eq:DFTperiodicity})
is that the DC $\tilde{x}_{0}$ and the Nyquist $\tilde{x}_{N/2}$ DFT
components are strictly real-valued: $\tilde{x}_{0}=\sum_{k=0}^{N-1}%
x_{k}=N\left\langle x\right\rangle $ and $\tilde{x}_{N/2}=\sum_{k=0}%
^{N-1}\left(  -1\right)  ^{k}~x_{k}$.

If $\mathbf{x}$ is a stochastic time-series, we may define the
auto-correlation matrix $\mathbf{C}$ with elements $C_{kk^{\prime}%
}=\left\langle x_{k}x_{k^{\prime}}\right\rangle $. For stationary processes,
the auto-correlation depends only on the lag $l=k^{\prime}-k,$ so that
$C_{l}\equiv C_{k,k+l}=\left\langle x_{k}x_{k+l}\right\rangle $. Then the
auto-correlation function has the following properties: $C_{0}=\left\langle
x_{k}x_{k}\right\rangle =\sigma^{2},$ $C_{l}=C_{-l}$, $\left\vert
C_{l}\right\vert \leq C_{0}=\sigma^{2}$, where $\sigma$ is the standard deviation.

We define the (two-sided) power spectral density (PSD) matrix $\mathbf{\tilde
{C}}$ with elements $\tilde{C}_{pp^{\prime}}\equiv\langle\tilde{x}_{p}\left(
\tilde{x}_{p^{\prime}}\right)  ^{\ast}\rangle$, which by
using\ (\ref{Eq:App:DFT-matrix}) can be related to the auto-correlation matrix
$\mathbf{C}$ as
\begin{align}
\mathbf{\tilde{C}} &  =N~\mathbf{UCU}^{\dagger}\,,\label{Eq:App:PSD-Auto-mat}%
\\
\mathbf{C} &  =N^{-1}\mathbf{U}^{\dagger}\mathbf{\tilde{C}U}\,.
\end{align}
For stationary processes, the PSD matrix is diagonal: $\tilde{C}_{pp^{\prime}%
}=\delta_{pp^{\prime}}\langle\left\vert \tilde{x}_{p}\right\vert ^{2}\rangle$.
We will simply refer to the diagonal matrix elements as $\tilde{C}_{p}%
\equiv\langle\left\vert \tilde{x}_{p}\right\vert ^{2}\rangle$. 
In particular,
\begin{align}
\tilde{C}_{p} &  =N\sum_{kk^{\prime}}U_{pk}C_{kk^{\prime}}U_{pk^{\prime}%
}^{\ast}= \nonumber\\
& \sum_{k,k^{\prime}=0}^{N-1}C_{kk^{\prime}}\exp\left(  i\frac{2\pi}%
{N}p\left(  k-k^{\prime}\right)  \right)  =\label{Eq:App:WKT}\\
&  =\sum_{l=-\left(  N-1\right)  }^{N-1}\left(  N-\left\vert l\right\vert
\right)  C_{l}\exp\left(  -i\frac{2\pi}{N}pl\right)   \approx \nonumber \\ 
& N\sum_{l=-\left(
N-1\right)  }^{N-1}C_{l}\exp\left(  -i\frac{2\pi}{N}pl\right)  .\nonumber
\end{align}
This is the statement of the Wiener-Khinchin theorem.

As an illustration of the Wiener-Khinchin theorem, consider a white-noise
process for which $C_{l}^{\mathrm{w.n.}}=\sigma^{2}\delta_{l,0}$. Then
Eq.~(\ref{Eq:App:WKT}) yields PSD $\tilde{C}_{p}^{\mathrm{w.n.}}=N\sigma^{2}$
regardless of the DFT frequency. This white-noise relation can be also derived
directly: $\tilde{C}_{p}^{\mathrm{w.n.}}=\langle\left\vert \tilde{x}%
_{p}\right\vert ^{2}\rangle=N\sum_{k}\sum_{k^{\prime}}U_{pk}C_{kk^{\prime}%
}^{\mathrm{w.n.}}U_{pk^{\prime}}^{\ast}=N\sigma^{2}\sum_{k}\sum_{k^{\prime}%
}U_{pk}\delta_{kk^{\prime}}U_{pk^{\prime}}^{\ast}=N\sigma^{2}%
(\underset{=\mathbf{I}}{\underbrace{\mathbf{UU}^{\dagger}})_{pp}}=N\sigma^{2}$.

We will need another property of PSDs: $\tilde{C}_{p}=\tilde{C}_{N-p}$. This
relation immediately follows from Eq.~(\ref{Eq:DFTperiodicity}).

\subsection{ Probability distribution in frequency space for deterministic
signal}

\label{Sec:App:Deterministic}

The derivation in this section essentially follows
Ref.~\cite{RomanoCornish2017}. It corrects errors in that work related to the
DC component and, in addition, includes the contribution from the Nyquist frequency.

Consider a data time series $\boldsymbol{d}$, which can contain both the
sought signal $\mathbf{s}$ and noise $\boldsymbol{n}$: $d_{k}=s_{k}+n_{k}$,
$k=\overline{0,N-1}$. The signal is prescribed by some model $M$. We assume
that the noise is Gaussian with $\left\langle n_{k}\right\rangle =0$, but not
necessarily white.\ Then the likelihood is given by the multi-variate Gaussian
distribution for residuals $\boldsymbol{n}=\boldsymbol{d}-\boldsymbol{s}$,
\begin{equation}
p\left(  \boldsymbol{d}|M,I\right)  =\frac{1}{\sqrt{\det\left(  2\pi
\boldsymbol{C}\right)  }}\exp\left(  -\frac{1}{2}\boldsymbol{n}^{\dagger
}\boldsymbol{C}^{-1}\boldsymbol{n}\right)  \,,
\end{equation}
where $I$ stands for the prior information proposition and the autocorrelation
matrix has elements $C_{kk^{\prime}}=\left\langle n_{k}n_{k^{\prime}%
}\right\rangle $. Now we rotate the argument of the exponential to the DFT
basis by inserting the identity $\mathbf{U}^{\dagger}\mathbf{U}=\mathbf{I}$:
$\boldsymbol{n}^{\dagger}\boldsymbol{C}^{-1}\boldsymbol{n}=\boldsymbol{n}%
^{\dagger}(\mathbf{U}^{\dagger}\mathbf{U})\boldsymbol{C}^{-1}(\mathbf{U}%
^{\dagger}\mathbf{U})\boldsymbol{n}=\left(  \mathbf{U}\boldsymbol{n}\right)
^{\dagger}\left(  \mathbf{U}\boldsymbol{C}\mathbf{U}^{\dagger}\right)
^{-1}\left(  \mathbf{U}\boldsymbol{n}\right)  $. The indicated groups can be
expressed in terms of DFT quantities: $\mathbf{U}\boldsymbol{n}=\tilde
{\boldsymbol{n}}/\sqrt{N}$, and $\mathbf{U}\boldsymbol{C}\mathbf{U}^{\dagger
}=N^{-1}\mathbf{\tilde{C}}$, see Eq.~(\ref{Eq:App:DFT-matrix}%
,\ref{Eq:App:PSD-Auto-mat}). Further, we use the fact that the PSD matrix is
diagonal, $\mathbf{\tilde{C}}=\mathrm{diag}(\tilde{C}_{p}),$ to invert the
matrix: $\mathbf{\tilde{C}}^{-1}=\mathrm{diag}\left(  \tilde{C}_{p}%
^{-1}\right)  $. Thus, the argument of the exponential is transformed into%

\begin{equation}
-\frac{1}{2}\boldsymbol{n}^{\dagger}\boldsymbol{C}^{-1}\boldsymbol{n}%
=-\frac{1}{2}\sum_{p=0}^{N-1}\frac{\left\vert \tilde{n}_{p}\right\vert ^{2}%
}{\tilde{C}_{p}},
\end{equation}
leading to the likelihood (we will recover the normalization factor below)
\begin{equation}
p_{\det}\left(  \boldsymbol{\tilde{d}}|M,I\right)  \propto\prod_{p=0}%
^{N-1}\exp\left(  -\frac{1}{2}\frac{\left\vert \tilde{n}_{p}\right\vert ^{2}%
}{\tilde{C}_{p}}\right)  .
\end{equation}
Here we used label \textquotedblleft det\textquotedblright\ to emphasize the
deterministic (as opposed to stochastic) nature of the signal. Apparently, the
spectral contributions become uncorrelated as a result of the basis rotation.
As the data are real-valued, we further use the redundancy $\left\vert
\tilde{n}_{N-p}\right\vert ^{2}=\left\vert \tilde{n}_{p}\right\vert
^{2},\tilde{C}_{N-p}=\tilde{C}_{p}$ and combine identical spectral
contributions (notice that $p=0$ and $p=N/2$ contributions are treated as
special cases as they do not have matching quantities).%

\begin{align}
p_{\det}\left(  \boldsymbol{\tilde{d}}|M,I\right)  \propto&\exp\left(
-\frac{1}{2}\left(  \frac{\left\vert \tilde{n}_{0}\right\vert ^{2}}{\tilde
{C}_{0}}+\frac{\left\vert \tilde{n}_{N/2}\right\vert ^{2}}{\tilde{C}_{N/2}%
}\right)  \right)  \times \nonumber \\
& \prod_{p=1}^{N/2-1}\exp\left(  -\frac{\left\vert \tilde
{n}_{p}\right\vert ^{2}}{\tilde{C}_{p}}\right)  .
\end{align}
The volume element associated with this probability distribution reads
\begin{equation}
d\operatorname{Re}\left(  \tilde{n}_{0}\right)  \,d\operatorname{Re}\left(
\tilde{n}_{N/2}\right)  \,\prod_{p=1}^{N/2-1}\,d\operatorname{Re}\left(
\tilde{n}_{p}\right)  d\operatorname{Im}\left(  \tilde{n}_{p}\right)  \,,
\end{equation}
where we took into account that the DC and the Nyquist values are strictly
real. The total number of random DFT variables is $N$ and it remains equal to
the number of sampled data points. To recover the normalization factor we can
compute the Jackobian $J$ of the transformation, so that $p\left(  \tilde
{D}|M,I\right)  =J\times p\left(  D|M,I\right)  .$ Alternatively, as the
transformation $\mathbf{U}$ is linear, we can simply normalize the
distribution. With $\int_{-\infty}^{+\infty}\exp\left[  -\frac{1}{2}x^{2}%
/\rho^{2}\right]  dx=\sqrt{2\pi\rho}$ and $\int_{-\infty}^{+\infty}\exp\left[
-(x^{2}+y^{2})/\rho^{2}\right]  dxdy=\pi\rho$, we arrive at the
frequency-domain likelihood%
\begin{equation}
p_{\det}\left(  \boldsymbol{\tilde{d}}|M,I\right)  =\prod_{p=0}^{N/2}\frac
{1}{(\beta_{p}^{-1}\pi\tilde{C}_{p})^{\beta_{p}}} e^{ -\beta_{p}%
\frac{\left\vert \tilde{n}_{p}\right\vert ^{2}}{\tilde{C}_{p}}}
\,.\label{Eq:App:DFTdistro}%
\end{equation}
Here $\beta_{j}=1$ except for the DC and the Nyquist components for which
$\beta_{0}=\beta_{N/2}=1/2$. It is worth emphasizing that we treated the
sought signal\ $\mathbf{s}$ as deterministic. In the above likelihood
$\tilde{n}_{p}\equiv$ $\tilde{d}_{p}-\tilde{s}_{p}$.

While we derived the DFT likelihood (\ref{Eq:App:DFTdistro}) rigorously, this
result is expected on qualitative grounds as the DFT components $\tilde{n}%
_{p}$ are sums of Gaussian random variables $n_{k}$, meaning that $\tilde
{n}_{p}$ is also Gaussian-distributed. Moreover, the DFT values are
uncorrelated, thereby the DFT likelihood (\ref{Eq:App:DFTdistro}) is a
product of individual Gaussians.

\subsection{Stochastic signal probability distribution and its likelihood in
frequency space}

\label{Sec:App:Stochastic}

Now let us assume that the sought signal $\mathbf{s}$ itself is stochastic and
Gaussian-distributed, with mean value being zero.  Then its probability
distribution (c.f., Eq.~(\ref{Eq:App:DFTdistro})) reads
\begin{equation}
p\left(  \boldsymbol{\tilde{s}}|M,I\right)  =\prod_{p=0}^{N/2}\frac{1}%
{(\beta_{p}^{-1}\pi\tilde{S}_{p})^{\beta_{p}}}\exp\left(  -\beta_{p}%
\frac{\left\vert \tilde{s}_{p}\right\vert ^{2}}{\tilde{S}_{p}}\right)  \,,
\end{equation}
where the signal PSD $\tilde{S}_{p}=\langle\left\vert \tilde{s}_{p}\right\vert
^{2}\rangle$. At the same time for a specific realization of the signal, the
likelihood is given by Eq.~(\ref{Eq:App:DFTdistro}),%
\begin{equation}
p\left(  \boldsymbol{\tilde{d}}|\tilde{S},I\right)  =\prod_{p=0}^{N/2}\frac
{1}{(\beta_{p}^{-1}\pi\tilde{C}_{p})^{\beta_{p}}} e^{  -\beta_{p}%
{\left\vert \tilde{d}_{p}-\tilde{s}_{p}\right\vert ^{2} / \tilde{C}_{p}%
} }  \,,
\end{equation}
where $\tilde{C}_{p}=\left\langle \left\vert \tilde{n}_{p}\right\vert
^{2}\right\rangle $ is the noise PSD. Then the likelihood for a stochastic
signal is obtained by marginalizing over realizations  of the signal,%
\begin{equation}
p_{\mathrm{stoh}}\left(  \boldsymbol{\tilde{d}}|M,I\right)  =\int
d\boldsymbol{\tilde{s}}~p\left(  \boldsymbol{\tilde{d}}|\tilde{S},I\right)
p\left(  \boldsymbol{\tilde{s}}|M,I\right)  .
\end{equation}
The integration can be carried out using explicitly, leading to%
\begin{equation}
p_{\mathrm{stoh}}\left(  \boldsymbol{\tilde{d}}|M,I\right)  =\prod_{p=0}%
^{N/2}\frac{1}{(\beta_{p}^{-1}\pi\tilde{\Sigma}_{p})^{\beta_{p}}} e^{
-\beta_{p}{\left\vert \tilde{d}_{p}\right\vert ^{2}/\tilde{\Sigma}_{p}%
} } \,,\label{Eq:App:DFTstoch}%
\end{equation}
with $\tilde{\Sigma}_{p}=\tilde{C}_{p}+\tilde{S}_{p}$. This result can be also
understood without explicit marginalization: consider a random variable
$\tilde{\eta}_{p}\equiv\tilde{s}_{p}+\tilde{n}_{p}=\tilde{d}_{p}$. Because
both $\tilde{s}_{p}$ and $\tilde{n}_{p}$ are Gaussian, their sum $\tilde{\eta
}_{p}$ is Gaussian as well and it is distributed according to  Eq.~(\ref{Eq:App:DFTdistro}).  The associated PSD is $\langle\left\vert
\tilde{\eta}_{p}\right\vert ^{2}\rangle=\left\langle \left(  \tilde{s}%
_{p}+\tilde{n}_{p}\right)  \left(  \tilde{s}_{p}+\tilde{n}_{p}\right)  ^{\ast
}\right\rangle =\langle\left\vert \tilde{s}_{p}\right\vert ^{2}\rangle
+2\operatorname{Re}\underbrace{\left\langle \tilde{s}_{p}\tilde{n}_{p}^{\ast
}\right\rangle }_{=0}+\langle\left\vert \tilde{n}_{p}\right\vert ^{2}%
\rangle=\tilde{C}_{p}+\tilde{S}_{p}=\tilde{\Sigma}_{p}$. Then the application
of Eq.~ (\ref{Eq:App:DFTdistro}) with $\tilde{n}_{p}\rightarrow\tilde{\eta}%
_{p}=\tilde{d}_{p}$ and$~\tilde{C}_{p}\rightarrow$ $\tilde{\Sigma}_{p}$
immediately yields Eq.~(\ref{Eq:App:DFTstoch}).


\subsection{Likelihoods for a network}

\label{Sec:App:Network}

This section generalizes frequency-space likelihoods for deterministic and
stochastic signals to a network of devices.

We consider a network of $N_{d}$ devices. The network nodes can be degenerate
and contain one or several devices. We will label the devices with letters at
the beginning of the alphabet $a,b,...$ and use superscripts for device labels
and subscripts for time stamps. Then time series $\left\{  x_{k}^{a}\right\}
$ refers to the time series of Sec.~\ref{Sec:App:DFT} for the $a$-th device.
Each device data time series $\left\{  d_{k}^{a}\right\}  $ can contain noise
component $\left\{  n_{k}^{a}\right\}  $, $\left\langle n_{k}^{a}\right\rangle
=0$. To streamline notation, we introduce super-vectors spanning the entire
network time series: $\underline{\boldsymbol{x}}=\left(  \left(  x_{1}%
^{1},x_{2}^{1},...,x_{N}^{1}\right)  ,\ldots,\left(  x_{1}^{N_{d}}%
,x_{2}^{N_{d}},...,x_{N}^{N_{d}}\right)  \right)  ^{T}$. Super-matrices
$\underline{\boldsymbol{C}}$ are defined in a similar fashion. We will use
underlined quantities to emphasize the use of this super-vector space.

Correlation matrix $\underline{\boldsymbol{C}}$ for a network is defined as a
matrix with elements $\underline{\boldsymbol{C}}_{kk^{\prime}}^{ab}%
\equiv\left\langle n_{k}^{a}n_{k^{\prime}}^{b}\right\rangle $. If devices are
uncorrelated, $\underline{\boldsymbol{C}}_{kk^{\prime}}^{ab}\equiv\delta
_{ab}\left\langle n_{k}^{a}n_{k^{\prime}}^{a}\right\rangle $. We will,
however, consider the general case, when the devices may share some noise
channel, introduced, for example, by the device comparison techniques such as
the use of inter-node optical fiber. The PSD matrix
$\underline{\boldsymbol{\tilde{C}}}$ is generalized to have elements
$\underline{\boldsymbol{\tilde{C}}}_{pp^{\prime}}^{ab}\equiv\langle\tilde
{n}_{p}^{a}\left(  \tilde{n}_{p^{\prime}}^{b}\right)  ^{\ast}\rangle.$

First, consider a deterministic signal $\left\{  s_{k}^{a}\right\}  $
prescribed by a model $M$. We suppose each datum is composed of the noise and
the sought signal: $d_{k}^{a}=s_{k}^{a}+n_{k}^{a}$. The time-domain likelihood
is given by the multi-variate Gaussian distribution%
\begin{equation}
p\left(  \underline{\boldsymbol{d}}|M,I\right)  =\frac{1}{\sqrt{\det\left(
2\pi\underline{\boldsymbol{C}}\right)  }}\exp\left(  -\frac{1}{2}%
\underline{\boldsymbol{n}}^{\dagger}\underline{\boldsymbol{C}}^{-1}%
\underline{\boldsymbol{n}}\right)  \,,
\end{equation}
with $I$ being prior information. Explicitly,%
\begin{equation}
\underline{\boldsymbol{n}}^{\dagger}\underline{\boldsymbol{C}}^{-1}%
\underline{\boldsymbol{n}}~\boldsymbol{=}\sum_{a,b=1}^{N_{d}}\sum
_{k,k^{\prime}=0}^{N-1}n_{k}^{a}\left(  \underline{\boldsymbol{C}}%
^{-1}\right)  _{kk^{\prime}}^{ab}n_{k^{\prime}}^{b}.
\end{equation}
Now we carry out the rotation to the DFT basis as in
Sec.~\ref{Sec:App:Deterministic}. The relevant DFT super-matrix
$\underline{\boldsymbol{U}}$ is block-diagonal and is composed of $N_{d}$ DFT
matrices $\boldsymbol{U}$. Then%
\begin{equation}
\underline{\boldsymbol{n}}^{\dagger}\underline{\boldsymbol{C}}^{-1}%
\underline{\boldsymbol{n}}~\boldsymbol{=}\frac{1}{N}%
\underline{\boldsymbol{\tilde{n}}}^{\dagger}\left(  \underline{\boldsymbol{U}%
}~\underline{\boldsymbol{C}}~\underline{\boldsymbol{U}}^{\dagger}\right)
^{-1}\underline{\boldsymbol{\tilde{n}}}=\underline{\boldsymbol{\tilde{n}}%
}^{\dagger}\underline{\boldsymbol{\tilde{C}}}^{-1}%
\underline{\boldsymbol{\tilde{n}}}\,.
\end{equation}
Here the DFT super-vectors are $\underline{\boldsymbol{\tilde{n}}}=\left(
\left(  \tilde{n}_{1}^{1},\tilde{n}_{2}^{1},...,\tilde{n}_{N}^{1}\right)
,...\left(  \tilde{n}_{1}^{N_{d}},\tilde{n}_{2}^{N_{d}},...,\tilde{n}%
_{N}^{N_{d}}\right)  \right)  ^{T}$, where the lower indexes enumerate DFT
frequencies.
To efficiently invert $\boldsymbol{\tilde{C}}$, we reshuffle the components of
$\underline{\boldsymbol{\tilde{n}}}$ so that $\underline{\boldsymbol{\tilde
{n}}}^{\prime}=\left(  \left(  \tilde{n}_{1}^{1},\tilde{n}_{1}^{2}%
,...,\tilde{n}_{1}^{N_{D}}\right)  ,...\left(  \tilde{n}_{N}^{1},\tilde{n}%
_{N}^{2},...,\tilde{n}_{N}^{N_{d}}\right)  \right)  ^{T}$, i.e., each
sub-group shares the same DFT frequency. Then the reshuffled PSD matrix is
block-diagonal, each block $\mathbf{\tilde{C}}_{p}$ corresponding to a given
DFT frequency: $\left(  \underline{\boldsymbol{\tilde{C}}}_{p}\right)
^{ab}=\langle\tilde{n}_{p}^{a}\left(  \tilde{n}_{p}^{b}\right)  ^{\ast}%
\rangle$. Indeed, for a stationary stochastic process, the network PSD is
diagonal in  frequency indexes for reasons discussed in Sec. \ref{Sec:App:DFT}%
: $\underline{\boldsymbol{\tilde{C}}}_{pp^{\prime}}^{ab}\equiv\delta
_{pp^{\prime}}\langle\tilde{n}_{p}^{a}\left(  \tilde{n}_{p}^{b}\right)
^{\ast}\rangle$. The reshuffling is useful because an inverse of a
block-diagonal matrix is itself a block-diagonal matrix, with sub-matrices
$\underline{\boldsymbol{\tilde{C}}}_{p}^{-1}$.

The resulting frequency-domain network likelihood for a deterministic signal
\begin{align}
p_{\det}&\left(  \underline{\boldsymbol{\tilde{d}}}|M,I\right)  = 
\prod
_{p=0}^{N/2}  \frac{1}{\left(  \beta_{p}^{-1}\pi\det\left(
\underline{\boldsymbol{\tilde{C}}}_{p}\right)  \right)  ^{\beta_{p}}} \times \label{Eq:App:Network-Deterministic}\\
& \exp\left(  -\beta_{p}\left(  \underline{\boldsymbol{\tilde{d}}}%
_{p}-\underline{\boldsymbol{\tilde{s}}}_{p}\right)  ^{\dagger}%
\underline{\boldsymbol{\tilde{C}}}_{p}^{-1}\left(
\underline{\boldsymbol{\tilde{d}}}_{p}-\underline{\boldsymbol{\tilde{s}}}%
_{p}\right)  \right)  \,, \nonumber
\end{align}
where data vectors $\underline{\boldsymbol{\tilde{d}}}_{p}\equiv\left(
\tilde{d}_{p}^{1},\tilde{d}_{p}^{2},\ldots,\tilde{d}_{p}^{N_{D}}\right)  $,
signal vectors $\underline{\boldsymbol{\tilde{s}}}_{p}\equiv\left(  \tilde
{s}_{p}^{1},\tilde{s}_{p}^{2},\ldots,\tilde{s}_{p}^{N_{D}}\right)  $, and DFTs
$\left(  \underline{\boldsymbol{\tilde{C}}}_{p}\right)  ^{ab}=\langle\tilde
{n}_{p}^{a}\left(  \tilde{n}_{p}^{b}\right)  ^{\ast}\rangle$. As in
Eq.~(\ref{Eq:App:DFTdistro}), $\beta_{p}=1$ except for the DC and the Nyquist
components for which $\beta_{0}=\beta_{N/2}=1/2$.

Following the steps of Sec.~\ref{Sec:App:Stochastic}, we generalize
Eq.~(\ref{Eq:App:Network-Deterministic}) to stochastic signals:
\begin{align}
p_{\mathrm{stoch}}\left(  \underline{\boldsymbol{\tilde{d}}}|M,I\right)
=\prod_{p=0}^{N/2}\frac{1}{\left(  \beta_{p}^{-1}\pi\det\left(
\underline{\boldsymbol{\tilde{\Sigma}}}_{p}\right)  \right)  ^{\beta_{p}}}%
e^{ -\beta_{p}~\underline{\boldsymbol{\tilde{d}}}_{p}^{\dagger
}\underline{\boldsymbol{\tilde{\Sigma}}}_{p}^{-1}\underline{\boldsymbol{\tilde
{d}}}_{p}}  \,,\label{Eq:App:Network-Stochastic}%
\end{align}
with $\underline{\boldsymbol{\tilde{\Sigma}}}_{p}%
=\underline{\boldsymbol{\tilde{C}}}_{p}+\underline{\boldsymbol{\tilde{S}}}%
_{p}$. The elements of the signal PSD matrix are $\left(
\underline{\boldsymbol{\tilde{S}}}_{p}\right)  ^{ab}=\langle\tilde{s}_{p}%
^{a}\left(  \tilde{s}_{p}^{b}\right)  ^{\ast}\rangle$. This result can be
obtained immediately following the arguments given just below
Eq.~(\ref{Eq:App:DFTstoch}). Alternatively, one could explicitly carry out
marginalization over signal values. In this case, the marginalization can be
aided by the convolution formula of two multi-variate Gaussian distributions,%
\begin{align}
\int  \frac{e^{-\frac{1}{2} \left(  \boldsymbol{x}-\boldsymbol{y}\right)
^{T}\boldsymbol{A}^{-1}\left(  \boldsymbol{x}-\boldsymbol{y}\right)  }}%
{\sqrt{\det\left(  2\pi\boldsymbol{A}\right)  }} &
\frac{e^{-\frac{1}%
{2}\boldsymbol{y}^{T}\boldsymbol{B}^{-1}\boldsymbol{y}}}{\sqrt{\det\left(
2\pi\boldsymbol{B}\right)  }}d\boldsymbol{y}= \nonumber\\
& \frac{e^{-\frac{1}{2}%
\boldsymbol{x}^{T}\left(  \boldsymbol{A}+\boldsymbol{B}\right)  ^{-1}%
\boldsymbol{x}}}{\sqrt{\det\left(  2\pi\left(  \boldsymbol{A}+\boldsymbol{B}%
\right)  \right)  }}.
\end{align}
More general convolution formula can be found in \cite{CornishRomano2013}. The
Eqs.~(\ref{Eq:App:Network-Deterministic},\ref{Eq:App:Network-Stochastic})
correct corresponding equations in Ref.~\cite{CornishRomano2013} for the DC and
Nyquist contributions.

\section{Estimators of the mean and variance of the coupling strength,
Eqs.~(\ref{Eq:gammaHat1},\ref{Eq:sigmaHat1})}

\label{Sec:App:CouplingStrength}

Here I derive the maximum-likelihood estimators of  of the mean and variance
of the coupling strength for stochastic signal. In the main text, the noise
PSD is defined as $\tilde{\rho}_{p}=\langle|\tilde{n_{p}}|^{2}\rangle$ and the
dark matter signal PSD as $\ \langle|\tilde{s_{p}}|^{2}\rangle=A^{2}\gamma
_{X}^{2}\langle|\tilde{\phi_{p}}|^{2}\rangle$, where $\gamma_{X}$ is the
coupling strength to dark matter and $A$ is the (fixed) device constant. We
will use the parameterization $\langle|\tilde{s_{p}}|^{2}\rangle=\Gamma
~\tilde{S}_{p}$, with $\Gamma\equiv\left(  \gamma_{X}\right)  ^{2}$ and
$\tilde{S}_{p}\equiv A^{2}\langle|\tilde{\phi_{p}}|^{2}\rangle$. 

The likelihood is given by Eq.~(\ref{Eq:App:DFTstoch}). For simplicity, we
assume that the DC component has been removed from the data and the Nyquist
component was filtered out. Then,%
\begin{equation}
p_{\mathrm{stoh}}\left(  \boldsymbol{\tilde{d}}|\Gamma,I\right)  =\prod
_{p=1}^{N/2-1}\frac{1}{\pi\left(  \tilde{\rho}_{p}+\Gamma~\tilde{S}%
_{p}\right)  } e^{ -\frac{\left\vert \tilde{d}_{p}\right\vert ^{2}%
}{\tilde{\rho}_{p}+\Gamma~\tilde{S}_{p}} }  \,.
\end{equation}
We would like to determine the posterior distribution $p\left(  \Gamma
|\boldsymbol{\tilde{d}},I\right)  $ for parameter $\Gamma$ given data
$\boldsymbol{\tilde{d}}|$  and prior information $I$. 
To this end we invoke the Bayes' theorem%
\begin{equation*}
p\left(  \boldsymbol{\tilde{d}},\Gamma|I\right)  =p\left(  \boldsymbol{\tilde
{d}}|I\right)  p\left(  \Gamma|\boldsymbol{\tilde{d}},I\right)  =p\left(
\Gamma|I\right)  p\left(  \boldsymbol{\tilde{d}}|\Gamma,I\right)  \,,
\end{equation*}
leading to%
\begin{equation}
p\left(  \Gamma|\boldsymbol{\tilde{d}},I\right)  =\frac{p\left(
\Gamma|I\right)  }{p\left(  \boldsymbol{\tilde{d}}|I\right)  }p_{\mathrm{stoh}%
}\left(  \boldsymbol{\tilde{d}}|\Gamma,I\right)  .
\end{equation}
Here $p\left(  \Gamma|I\right)$ gives the prior probability density for the parameter $\Gamma$. We assume that there is no prior
knowledge about the value of the parameter. For example, we could take
$p\left(  \Gamma|I\right)$ to be a uniform distribution in the range
$\left[  0,\Gamma_{\max}\right]$, where $\Gamma_{\max}$ is sufficiently
large.  $p\left(  \boldsymbol{\tilde{d}}|I\right)$ is a normalization constant.

Now we bring the posterior distribution $p\left(  \Gamma|\boldsymbol{\tilde
{d}},I\right)  $ into a Gaussian form (strictly speaking, we are using the
Laplace approximation),
\begin{equation}
p\left(  \Gamma|\boldsymbol{\tilde{d}},I\right)  \approx\mathcal{C}\exp\left(
-\frac{\left(  \Gamma-\hat{\Gamma}\right)  ^{2}}{2\left(  \hat{\sigma}%
_{\Gamma}\right)  ^{2}}\right)  ,~0<\Gamma<\Gamma_{\max},
\end{equation}
where $\hat{\Gamma}$ and $\hat{\sigma}_{\Gamma}$ are the maximum likelihood
estimators for the mean and the standard deviation and are to be determined. The
proportionality factor $\mathcal{C}$ does not depend on $\Gamma$. The
probability maximum is reached at $\Gamma=\hat{\Gamma}$. Consider an auxiliary construction,
\begin{equation}
\ln\left(  p\left(  \Gamma|\boldsymbol{\tilde{d}},I\right)  \right)
\approx\ln\mathcal{C}-\frac{\left(  \Gamma-\hat{\Gamma}\right)  ^{2}}{2\left(
\hat{\sigma}_{\Gamma}\right)  ^{2}}.
\end{equation}
Then the estimator values can be determined from%
\begin{align}
& \left[  \frac{d}{d\Gamma}\ln\left(  p\left(  \Gamma|\boldsymbol{\tilde{d}%
},I\right)  \right)  \right]  _{\Gamma=\hat{\Gamma}}=0,\\
& \frac{1}{\left(  \hat{\sigma}_{\Gamma}\right)  ^{2}}=-\left[  \frac{d^{2}%
}{d^{2}\Gamma}\ln\left(  p\left(  \Gamma|\boldsymbol{\tilde{d}},I\right)
\right)  \right]  _{\Gamma=\hat{\Gamma}},
\end{align}
where $p\left(  \Gamma|\boldsymbol{\tilde{d}},I\right)  $ is to be replaced
with $~p_{\mathrm{stoh}}\left(  \boldsymbol{\tilde{d}}|\Gamma,I\right)  $,
Eq.~(\ref{Eq:App:DFTstoch}). Evaluating the derivatives and linearizing the
resulting expressions for weak signals, $\Gamma\,\tilde{S}_{p}\ll\tilde{\rho
}_{p}$, we arrive at
\begin{align}
& \frac{1}{\left(  \hat{\sigma}_{\Gamma}\right)  ^{2}}=\sum_{p=1}%
^{N/2-1}\left(  2\left\vert \tilde{d}_{p}\right\vert ^{2}/\tilde{\rho}%
_{p}-1\right)  \left(  \frac{\tilde{S}_{p}}{\tilde{\rho}_{p}}\right)  ^{2},\\
& \hat{\Gamma}=\left(  \hat{\sigma}_{\Gamma}\right)  ^{2}\sum_{p=1}%
^{N/2-1}\left(  \left\vert \tilde{d}_{p}\right\vert ^{2}/\tilde{\rho}%
_{p}-1\right)  \frac{\tilde{S}_{p}}{\tilde{\rho}_{p}} \,.
\end{align}
With $\hat{\gamma}_{x}=\sqrt{\hat
{\Gamma}}$ and $\hat{\sigma}_{\gamma_{X}}=\sqrt{\hat{\sigma}_{\Gamma}}$,
these equations lead to Eqs.~(\ref{Eq:gammaHat1},\ref{Eq:sigmaHat1}) of the main text.

\section{Derivation of Eq.~(\ref{Eq:sigma1Simp}) }
\label{Sec:App:SumToIntegral}

We start with the estimate for the standard deviation of the coupling strength%
\begin{equation}
\sigma_{\gamma_{X}}^{(1)}\approx A^{-1}\left\{  \sum_{p=1}^{N/2-1}\left(
\frac{\langle|\tilde{\phi_{p}}|^{2}\rangle}{\tilde{\rho}_{p}}\right)
^{2}\right\}  ^{-1/4}\,.
\end{equation}
For white noise $\tilde{\rho}_{p}=N\sigma^{2}.$%
\begin{equation}
\sigma_{\gamma_{X}}^{(1)}\approx A^{-1}\sigma\left\{  \frac{1}{N^{2}}%
\sum_{p=1}^{N/2-1}\langle|\tilde{\phi_{p}}|^{2}\rangle^{2}\right\}  ^{-1/4}\,.
\end{equation}
Further, we use the relationship of the discretized field PSD to the dark
matter line shape
\begin{equation}
\langle|\tilde{\phi_{p}}|^{2}\rangle=\frac{\pi N}{\Delta_{t}}\Phi_{0}%
^{2}F\left(  \omega_{p}\right)  \,,
\end{equation}
leading to
\begin{align}
\sigma_{\gamma_{X}}^{(1)} &  \approx A^{-1}\sigma\left\{  \frac{1}{N^{2}}%
\sum_{p=1}^{N/2-1}\frac{\pi^{2}N^{2}}{\Delta_{t}^{2}}\Phi_{0}^{4}\left(
F\left(  \omega_{p}\right)  \,\right)  ^{2}\right\}  ^{-1/4}=\nonumber\\
&  \,\left(  A\Phi_{0}\right)  ^{-1}\sigma\left\{  \frac{\pi^{2}}{\Delta
_{t}^{2}}\sum_{p=1}^{N/2-1}\left(  F\left(  \omega_{p}\right)  \,\right)
^{2}\right\}  ^{-1/4}.
\end{align}
Now we can convert the sum to an integral by introducing the DFT frequency step
$\Delta_{\omega}=\frac{2\pi}{N\Delta_{t}}$:
\begin{align}
\sigma_{\gamma_{X}}^{(1)} &  =\left(  A\Phi_{0}\right)  ^{-1}\sigma\left\{
\frac{\pi N }{2 \Delta_{t}}\sum_{p=1}^{N/2-1}%
\Delta_{\omega}\left(  F\left(  \omega_{p}\right)  \,\right)  ^{2}\right\}
^{-1/4}=\nonumber\\
&  \left(  A\Phi_{0}\right)  ^{-1}\sigma\left\{  \frac{\pi^{2}}{\Delta_{t}%
}\frac{N}{2\pi}\int_{0}^{\infty}d\omega~\left(  F\left(  \omega_{p}\right)
\,\right)  ^{2}\right\}  ^{-1/4}.
\end{align}
For the fiducial value of $\eta=1$,
\begin{align*}
&\int_{0}^{\infty}d\omega \, \left(  F\left(  \omega_{p}\right)  \,\right)  ^{2}
  = \tau_{c}^{2}~\frac{1}{2\pi e^{2}} \times \\
&\int_{\omega_{\phi}^{\prime}-\left(2\tau_{c}\right)^{-1}  }^{\infty}e^{-2\left(  \omega-\omega_{\phi}^{\prime}\right)
\tau_{c}}\sinh^{2}\left(  \sqrt{1+2\left(  \omega-\omega_{\phi}^{\prime
}\right)  \tau_{c}}\right)  =\nonumber\\
&  =\tau_{c}~\frac{1}{4\pi e^{2}}\int_{-1}^{\infty}e^{-x}\sinh^{2}\left(
\sqrt{1+x}\right)  =\frac{\operatorname{erf}\left(  1\right)  }{8\sqrt{\pi}%
}\tau_{c}.
\end{align*}
Thus,
\begin{align}
\sigma_{\gamma_{X}}^{(1)} &  =\left(  A\Phi_{0}\right)  ^{-1}\sigma\left\{
\frac{\sqrt{\pi}\operatorname{erf}\left(  1\right)  }{16}\right\}
^{-1/4}\left\{  N\times\left(  \frac{\tau_{c}}{\Delta_{t}}\right)  \right\}
^{-1/4}\nonumber\\
&  \approx2.4~\left(  \frac{\Delta_{t}}{N\tau_{c}}\right)  ^{1/4}\frac{\sigma
}{A\Phi_{0}} \, ,
\end{align}
which is Eq.~(\ref{Eq:sigma1Simp}) of the main text.

%
%

\end{document}